\documentclass[12pt]{article}
\usepackage[dvips]{graphics,color,epsfig}
\input{babarsym.tex}
\begin{document}


%
%

\begin{center}

{\Large{$B$ physics at the D\O\ Experiment at Fermilab}}
\vspace{0.5 in}

Vivek Jain \\
Department of Physics, State University of New York, \\
Albany, NY

\end{center}



\begin{abstract}

We discuss recent $B$ physics results from the D\O\ experiment at
Fermilab\footnote{Invited review article to appear in Modern Physics
Letters A.}. The results presented here use data sets with integrated
luminosities ranging from $\sim 200-440 $ pb$^{-1}$, collected at the
Tevatron, between April 2002 and August 2004, at a center of mass
energy of $p {\bar p}$ collisions of 1.96 TeV.

\end{abstract}

{PACS Nos.: 13.20.He, 13.25.Hw, 14.20.Mr, 14.40.Nd, 14.40.Gx}

\section{Introduction}	

An understanding of flavour dynamics is crucial since any unified
theory will have to account for the presence of three left-handed
families, measured mixing angles and masses of various quarks and
hadrons, {\it etc.}  The study of bottom hadrons provides unique
insights into the nature of the weak as well as the strong
interaction, and also provides a window into beyond-Standard Model (BSM)
effects\cite{Btheory}.

The study of bottom hadrons at the Fermilab Tevatron ($p{\bar p}$
collisions at a center of mass energy of 1.96 TeV) has many advantages
over that at the currently operating $e^+e^-$ ``$B$-factories'' at SLAC
and KEK, {\it viz.}, the production cross-section of $b{\bar b}$
quarks is about 150,000 times larger, and all species of bottom
hadrons are produced.\footnote{Although all $B$-hadron species were also
produced at LEP, $\sigma_{b \bar b}$ at the Tevatron is about 20,000
times larger.}  However, experimental conditions are not as clean; for
instance, the total $p{\bar p}$ cross-section is more than two orders
of magnitude larger than that for $b{\bar b}$ production. This implies
that experiments at the Tevatron are crucially dependent on designing
appropriate triggers; the collision rate at D\O\ is about 2.5 MHz,
whereas the experiment can only write out data at about 50-100 Hz.

The $B$ physics program at D\O\, is designed to be complementary to
the program at the $B$-factories at SLAC and KEK and includes studies
of \Bs oscillations, searches for rare decays such as $B_s \rightarrow
\mu^+ \mu^-$, $B$ spectroscopy, {\it e.g.,} $B^*_J$, lifetimes of
$B$ hadrons, search for the lifetime difference in the $B_s$ CP
eigenstates, study of beauty baryons, $B_c$ mesons, quarkonia (\jpsi,
$\chi_c$, $\Upsilon$), b production cross-section, etc.

One of the more important topics in $B$ physics is the search for
\BsBsb mixing.  Global fits, assuming that the CKM matrix is unitary
and the Standard Model (SM) is correct, indicate that the 95\% CL
interval\cite{CERN:yellow} for the mixing frequency, $\Delta M_s$, is
[14.2-28.1] ps$^{-1}$. The current limit\footnote{The limit is derived
by combining limits from 13 different measurements.} is $\Delta M_s >
14.9 {\rm ps}^{-1}$ at the 95\% CL\cite{CERN:yellow}. A measured value
of $\Delta M_s$ much larger than the upper limits given here could
imply new physics contributing to the box diagrams, e.g., extra Higgs
bosons or squarks and/or gluinos\cite{Nierste}.

\section{D\O\ detector}

For the current run of the Tevatron (Run II), the D\O\ detector went
through a major upgrade\cite{D0det}. The inner tracking system was
completely revamped. The detector now includes a Silicon tracker,
surrounded by a Scintillating Fiber tracker, both of which are
immersed in a 2 Tesla solenoidal magnetic field. Pre-shower counters,
to aid in electron and photon identification are located before the
calorimeter. The muon system has also been improved, {\it e.g.} more
shielding was added to reduce beam background. New trigger and data
acquisition systems were also installed.

The D\O\ detector has excellent tracking and lepton acceptance. Tracks
with pseudo-rapidity ($\eta$) as large as 2.5 ($\theta \sim
10^{\circ}$) and transverse momentum ($p_T$) as low as 180 MeV/c are
reconstructed.

The muon system can identify muons within $|\eta| < 2.0$. The minimum
p$_T$ of the reconstructed muons varies as a function of $\eta$. In
most of the results presented here, we required muons to have $p_T >
2$ GeV/c. Low pT electron identification is being worked on, currently
we are limited to $p_T > 2$ GeV/c and $|\eta| < 1.1$; however, we are
working to improve both the momentum and $\eta$ coverage.

D\O\ employs a three level trigger. Triggers at Level 1, which are
formed by individual detector sub-systems, and Level 2, where they are
further refined, are constructed using custom hardware. At Level 3,
the entire event is read out by a farm of computers which perform a
simplified event reconstruction to further refine the selection
criteria for interesting events. Currently, the input rate to Level 1
is 2.5 MHz and the output is 1600 Hz; the output rates for Level 2 and
3 are 800 Hz and 60 Hz respectively. Improvements are foreseen that
could improve the output rates to 2-2.5 kHz, 1.2 kHz and 100 Hz for
Levels 1, 2 and 3, respectively.

\section{Data Sample}

The results presented here are based on data collected between April
2002 and August 2004. The data correspond to an integrated luminosity
of about 440 pb$^{-1}$, however the analyses presented here used
anywhere from 200-440 pb$^{-1}$.  Events enriched in $B$ hadrons were
collected with a dimuon and single muon triggers. To reduce the data
rate, a luminosity dependent prescale was applied to the single muon
trigger (the prescale was 1 for instantaneous ${\cal L}
<20\times10^{30} {\rm cm}^{-2} {\rm s}^{-1}$). Both triggers require
that muons have hits in all layers of the muon system which implies
that they have total momentum $ \ge 3$ GeV.

Many of these results have been accepted for publication. Details on
the other analyses can be found on the D\O\ $B$ physics group web
page\cite{D0Bweb}.

\section{New Particles}

We first review results dealing with particles which either have never
been seen before or have been recently observed, {\it viz.}, $X(3872),
B^*_J(5732)$ and $B_c$.

\subsection{X(3872)}

\begin{figure}[pt]
\hspace{0.5cm} 
\parbox{6.5in}{
\hfil
\begin{minipage}[b]{6.5cm}
\centerline{\psfig{file=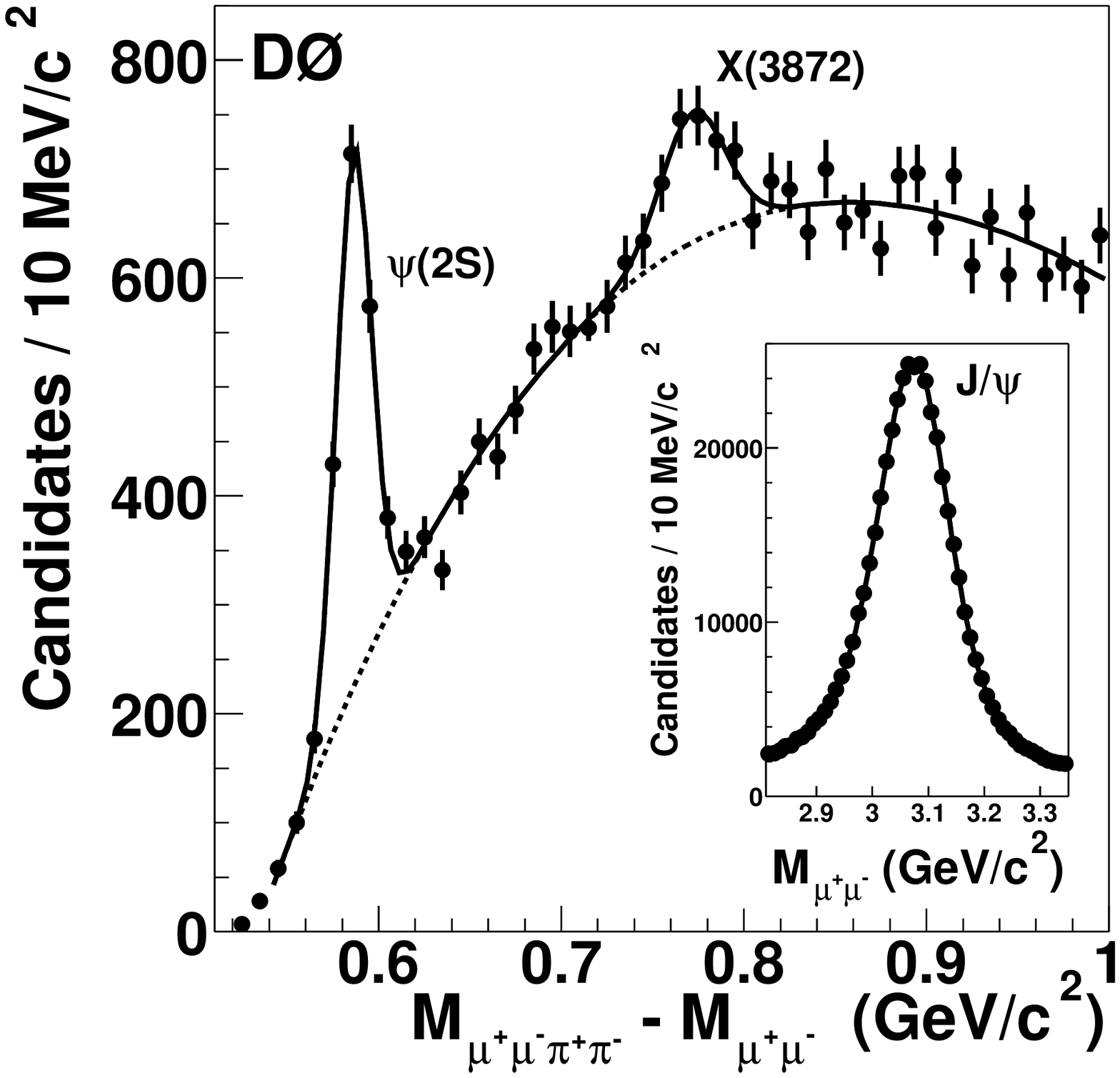,width=7cm}}
\vspace*{6pt}
\centerline{\footnotesize (a)}
\end{minipage}
\qquad\qquad\qquad
\begin{minipage}[b]{6.5cm}
\centerline{\psfig{file=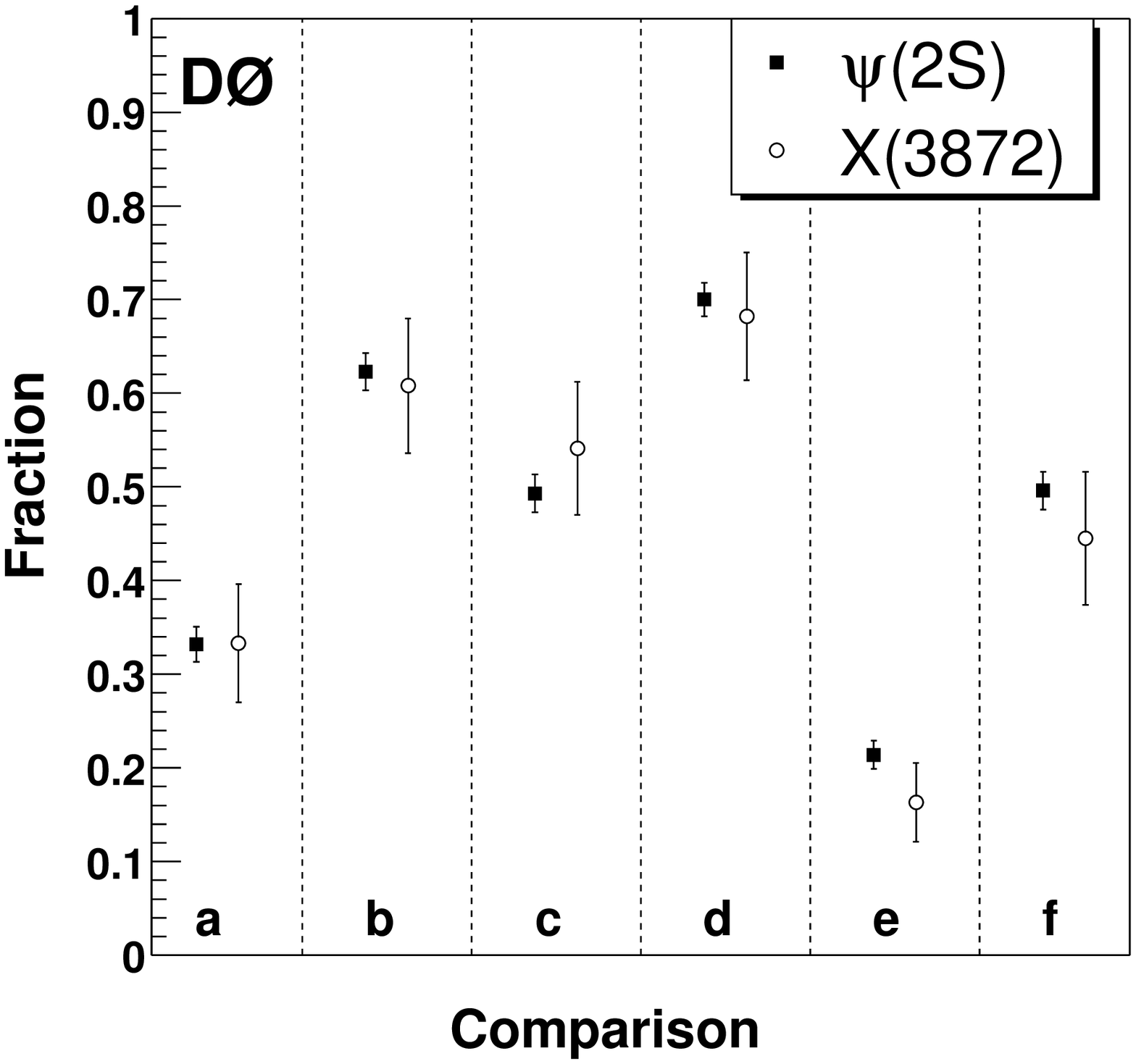,width=7cm}}
\vspace*{6pt}
\centerline{\footnotesize (b)}
\end{minipage}
}
\caption{(a) $\Delta M = M(\mu^{+} \mu^{-} \pi^{+} \pi^{-}) -
M(\mu^{+}\mu^{-})$ after all cuts. In the inset, we show the $J/\psi
\rightarrow \mu^+ \mu^-$ mass spectrum.  (b) Comparison of various
production and decay variables for the X(3872) and $\psi(2S)$ - see
text for details. \label{D0X}}
\end{figure}

X(3872) was first observed by the Belle collaboration\cite{OldX} via
$B\rightarrow X(3872)K, X \rightarrow J/\psi \pi^+ \pi^-$, and was
subsequently observed (inclusively) by the CDF
collaboration\cite{OldX} in $p {\bar p}$ collisions (in the same
$X(3872)$ final state).  In Fig.~\ref{D0X}(a), we present evidence for
$X(3872)\rightarrow J/\psi \pi^+\pi^-$, observed by the D\O\
collaboration\cite{D0Xpaper}; the inset shows the mass distribution of
$J/\psi \rightarrow \mu^+\mu^-$ candidates.  To improve resolution,
the mass difference $\Delta M = M(\mu^{+} \mu^{-} \pi^{+} \pi^{-}) -
M(\mu^{+}\mu^{-})$ is used. We observe $522\pm100$ $X(3872)$
candidates and measure $\Delta M = 774.9 \pm 3.1 (stat) \pm 3.0
(syst)$ MeV/c$^2$; this value is consistent with other measurements of
$M(X(3872))$ and is based on an integrated luminosity of 230
pb$^{-1}$.

It is not known whether the X(3872) is a normal charmonium state or
something more exotic, {\it e.g.,} $D{\bar D}$ molecule or $ccg$
hybrid\cite{Quigg}. To understand its nature we compare some of its
production and decay properties with the well-known charmonium state
$\psi(2S)$\cite{PDG}, which can also be seen in Fig~\ref{D0X}(a). In
Fig.~\ref{D0X}(b), we present the fraction of X(3872) ($\psi(2S)$)
candidates which satisfy certain criteria, (a) $p_T>15$ GeV/c, (b)
rapidity, $|y|<1$, (c) helicity of the $\pi\pi$ system,
$|cos(\theta_{\pi})|<0.4$, (d) effective proper decay length $<0.01$
cm, (e) Isolation\footnote{Isolation is defined as the ratio of the
$p_T$ of the $X(3872)$ to that of the $X(3872)$ and all other
particles within a cone of 0.7} = 1 and (f) helicity of the $\mu\mu$
system, $|cos(\theta_{\mu})|<0.4$. In all these variables, the X(3872)
appears to behave like the $\psi(2S)$.

Other studies include searching for the charged partner, {\it e.g.,}
$X^+ \rightarrow J/\psi \pi^+ \pi^0$ or for radiative decays like
$X(3872) \rightarrow \chi_c \gamma$; a signal for the former would
rule out the charmonium hypothesis whereas a signal for the latter
would confirm it as a charmonium state.

\subsection{P-wave mesons: $B^*_J(5732)$}

Hadrons which contain one heavy quark, $m_Q \gg \Lambda_{QCD}$, are
subject to additional QCD symmetries. As $m_Q \rightarrow \infty$, the
heavy quark decouples and the properties of the hadron are given by
light degrees of freedom({\it Ldof}), {\it i.e.,} light quark(s) and
gluons; this is known as Heavy Quark Symmetry (HQS)\cite{Isgur}. In
this limit, mesons belong to degenerate doublets given by $J^P = (j_l
\pm 1/2)^{\pi_l}$, where $J,j_l$ are the total angular momenta of the
meson and the {\it Ldof}, and $P, \pi_l$ are their respective
parities; also, $j_l = s_l + L$, where $s_l$ is the spin of the {\it
Ldof} and $L$ is the angular momentum between the {\it Ldof} and the
heavy quark. Degeneracy is broken due to finite $m_Q$ and so the
effect is larger for charm mesons than bottom mesons.

$L=0$ gives one doublet with $j_l = \frac{1}{2}$ and $J=0,1$ which
corresponds to the well-measured $B, B^*$ mesons\cite{PDG}. For $L=1$,
we get two doublets denoted as $j_l = \frac{1}{2}, J=0,1$ and
$j_l=\frac{3}{2}, J=1,2$; collectively, these four mesons are referred
to as $B^*_J$ mesons. Angular momentum and parity conservation
constrains the strong decays of the $L=1$ mesons, with the $J=0$ state
expected to decay to $B\pi$, $J=1$ states to $B^*\pi$ and $J=2$ state
to $B\pi$ and $B^*\pi$; $j_l = \frac{1}{2}$ decay via S-wave and are
expected to be broad while $j_l = \frac{3}{2}$ states decay via D-wave
and are expected to be narrow.

In the case of (non-strange) charm mesons all four $L=1$ states have
been observed, and their behaviour agrees with
expectations\cite{PDG,Belledouble}. On the other hand, the case
of $L=1$ charm-strange mesons states is more interesting. The $j_l =
\frac{3}{2}$ doublet decays to the favoured $D^{(*)}K$ final
states\cite{PDG}, whereas the $j_l=\frac{1}{2}$ states were lighter
than expected and could only decay to $D_s^{(*)}\pi^0$ final
states\cite{Newdouble}; this also causes them to be narrower than
expected\cite{Eichtendouble}.

\begin{figure}[hbt]
\hspace{0.5cm} 
\parbox{6.5in}{
\hfil
\begin{minipage}[b]{6.5cm}
\centerline{\psfig{file=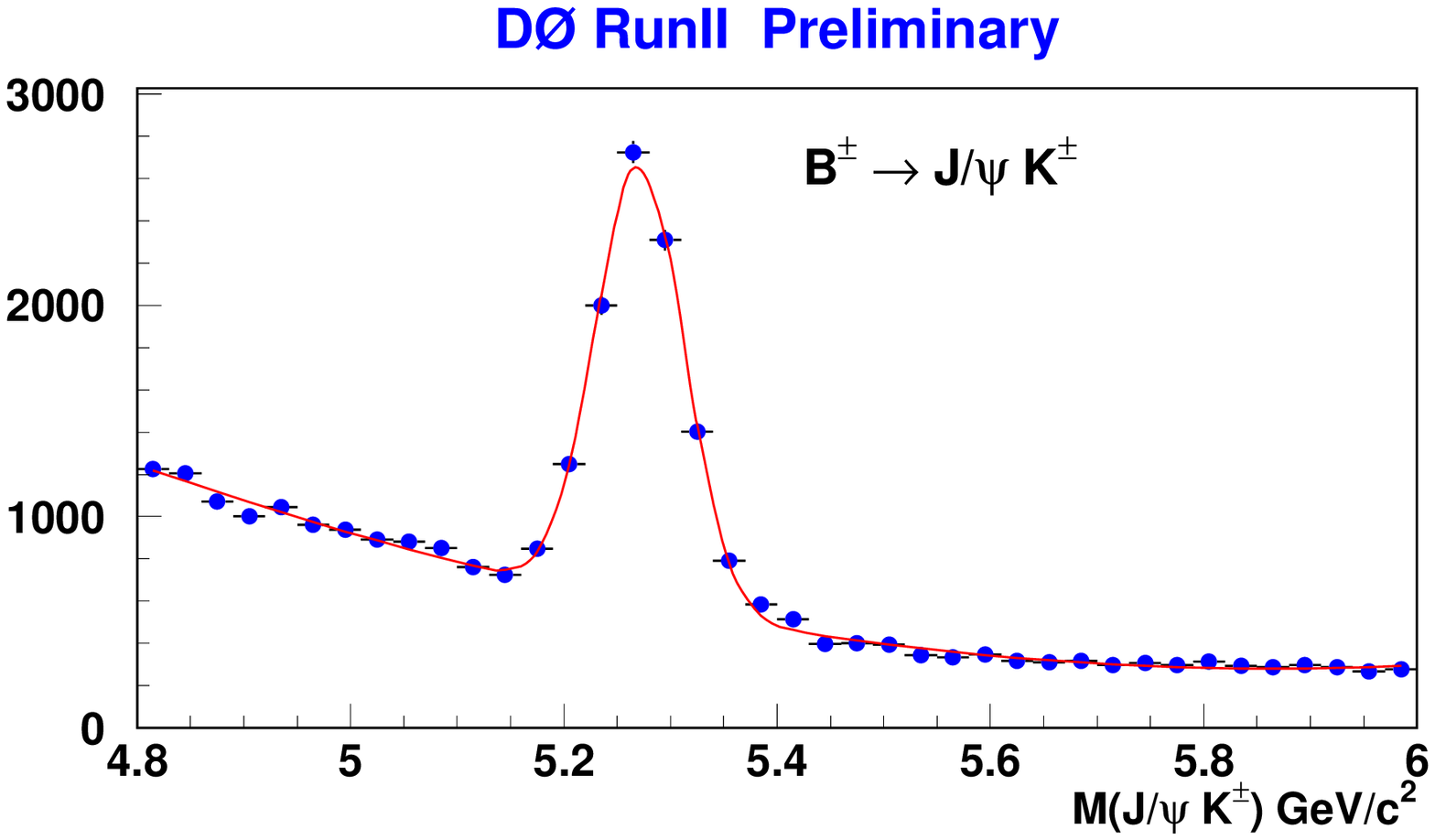,width=7cm}}
\vspace*{6pt}
\centerline{\footnotesize (a)}
\end{minipage}
\qquad\qquad\qquad
\begin{minipage}[b]{6.5cm}
\centerline{\psfig{file=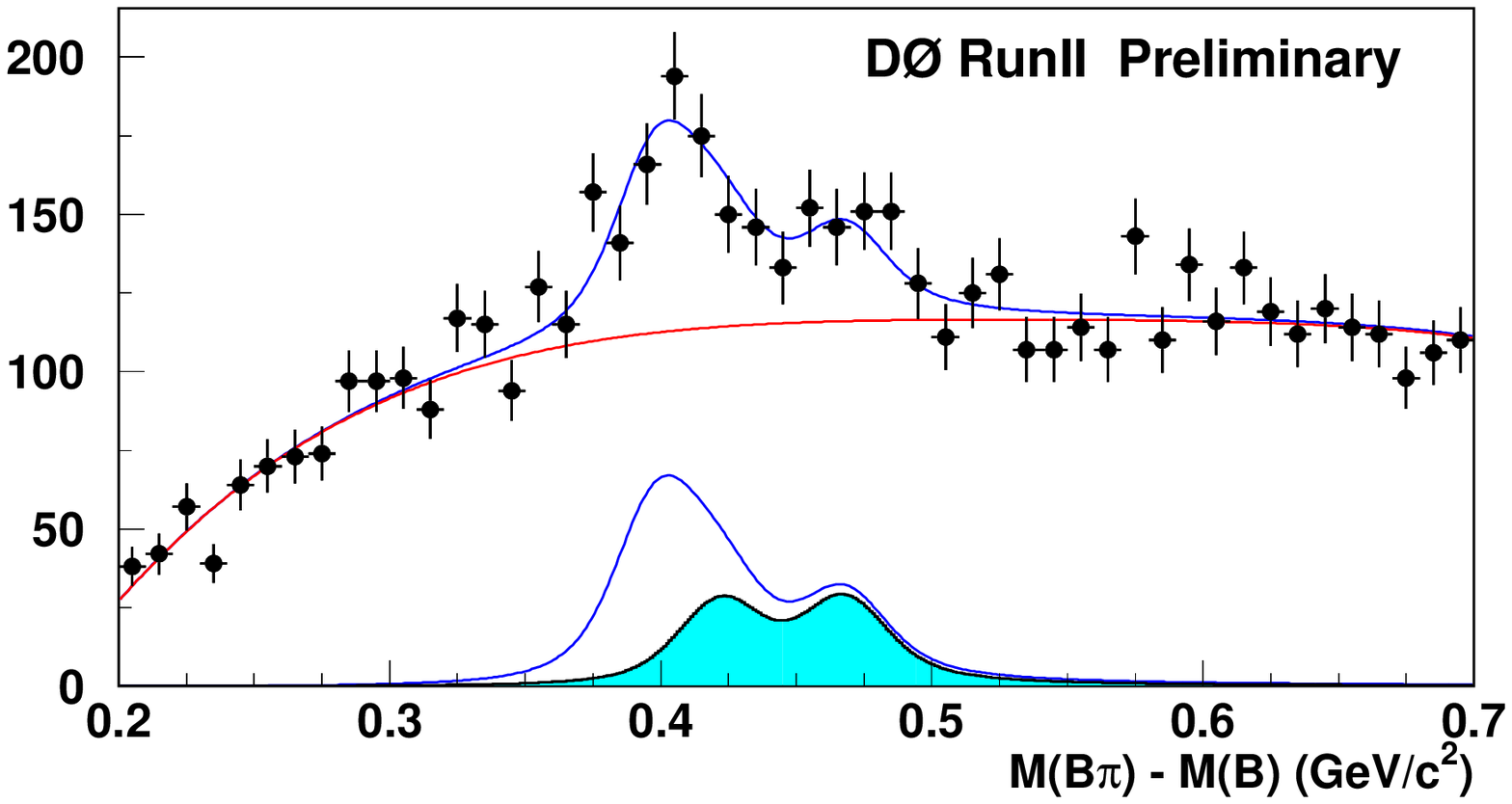,width=7cm}}
\vspace*{6pt}
\centerline{\footnotesize (b)}
\end{minipage}
}
\caption{Mass distribution for (a) $B^+ \rightarrow J/\psi K^+$, (b)
B$^{**}$ presented as $\Delta M = M(B\pi) - M(B)$ \label{D0Bplots}}
\end{figure}

In the case of $B^*_J$, an excess of events has been observed with
an average mass of $5698\pm8$ MeV/c$^2$, although the four $L=1$
states have not been individually observed\cite{PDG}; in addition,
only one of the experiments exclusively reconstructed the $B$
meson. 

We use our large sample of fully reconstructed $B$ mesons, (a) $B^+
\rightarrow J/\psi K^+$ (Fig~\ref{D0Bplots}(a)), (b) $B^0 \rightarrow
J/\psi K^{*0}$ and (c) $B^0 \rightarrow J/\psi K_s^0$ to search for
the narrow $B^*_J$ mesons; in $\sim 350$ pb$^{-1}$ of data, we have
the following yields, (a) $7217\pm127$, (b) $2826\pm93$ and (c)
$624\pm41$. Since the mass difference between $B^{*+}_J$ and
$B^{*0}_J$ is expected to be small, we add all three $B$ modes and
perform a combined search. In addition, to reduce resolution effects,
we plot the mass difference, $\Delta M = M(B^{+,0}\pi^{\pm}) -
M(B^{+,0})$, where the $\pi^{\pm}$ is required to be consistent with
coming from the primary vertex.

The mass difference plot, as shown in Fig.~\ref{D0Bplots}(b) has a
structure consistent with three components which correspond to the two
states, $B_1$ and $B_2^*$ (which make up the $j_l = \frac{3}{2}$
doublet), (a) $B_1 \rightarrow B^* \pi^{\pm}$, (b) $B_2^* \rightarrow
B^* \pi^{\pm}$ and (c) $B_2^* \rightarrow B \pi^{\pm}$.  The $B^*$
decays to a $B\gamma$ final state and the soft photon (in the
c.m. $E_{\gamma} \sim 46$ MeV) is not observed. This causes $\Delta M$
from the (a) and (b) to be shifted down by 46 MeV while it is in the
correct place for (c). Allowing for all three of these sources in the
final fit, we observe a total of $536\pm114$ events. Using input from
HQS, we assign $273\pm59$ to (a) and $131\pm30$ events each to (b) and
(c). In addition, we measure $M(B_1) = 5724\pm4 (stat) \pm7(syst)$
MeV/c$^2$, $M(B_2^*)-M(B_1) = 23.6\pm7.7\pm3.9$ MeV/c$^2$, and
assuming that $\Gamma_2 = \Gamma_1$ we obtain $23\pm12\pm9$ MeV for
the natural widths, in agreement with theoretical
expectations\cite{Theorydouble}. This is the first observation of the
narrow $B^*_J$ states.

Future studies will include separate fits for $B^{*+}_J$ and
$B^{*0}_J$ , measuring the production rate of $L=1 \: B$ mesons
relative to $L=0\: B$ mesons, measurement of the spin-parity of these
states, and a search for $B^*_{sJ}$.

\subsection{$B_c$ meson}

The $B_c$ meson consisting of a $b$ and a $c$ quark is the heaviest of
the flavoured ground state mesons that can exist; the top quark decays
before it can hadronize into a meson. Since it consists of two heavy
quarks, theoretical tools used to describe $c{\bar c}$ and $b{\bar b}$
mesons can be employed in its study\cite{Bctheory}. The $B_c$ meson
has non-zero flavour and thus it only has weak decays. It has been
observed by the CDF collaboration\cite{CDFBc}; the yield was
$20.4^{+6.2}_{-5.5}$ events in $110$ pb$^{-1}$ of $p {\bar p}$
collisions. The lifetime is expected to be closer to charm hadron
lifetimes ($\le 1$ ps) than to those of other beauty hadrons ($\sim
1.5 ps$).

A particularly attractive mode for its observation at D\O\ is the
semileptonic decay, $B_c \rightarrow J/\psi \mu \nu X$; the presence
of three muons in the final state makes it easy to trigger on. In
addition, backgrounds (branching fraction) are expected to be lower
(higher) than for the exclusive mode, $J/\psi\pi$. However, due to
missing particles in the final state, {\it e.g.,} neutrino (and maybe
pions), the determination of the mass and lifetime has to rely on
Monte Carlo simulations based on the ISGW model\cite{ISGW}. Systematic
effects are studied by using different decay models, {\it e.g.,} V-A,
{\it etc}.

\begin{figure}[hbt]
\hspace{0.5cm} 
\parbox{6.5in}{
\hfil
\begin{minipage}[b]{6.5cm}
\centerline{\psfig{file=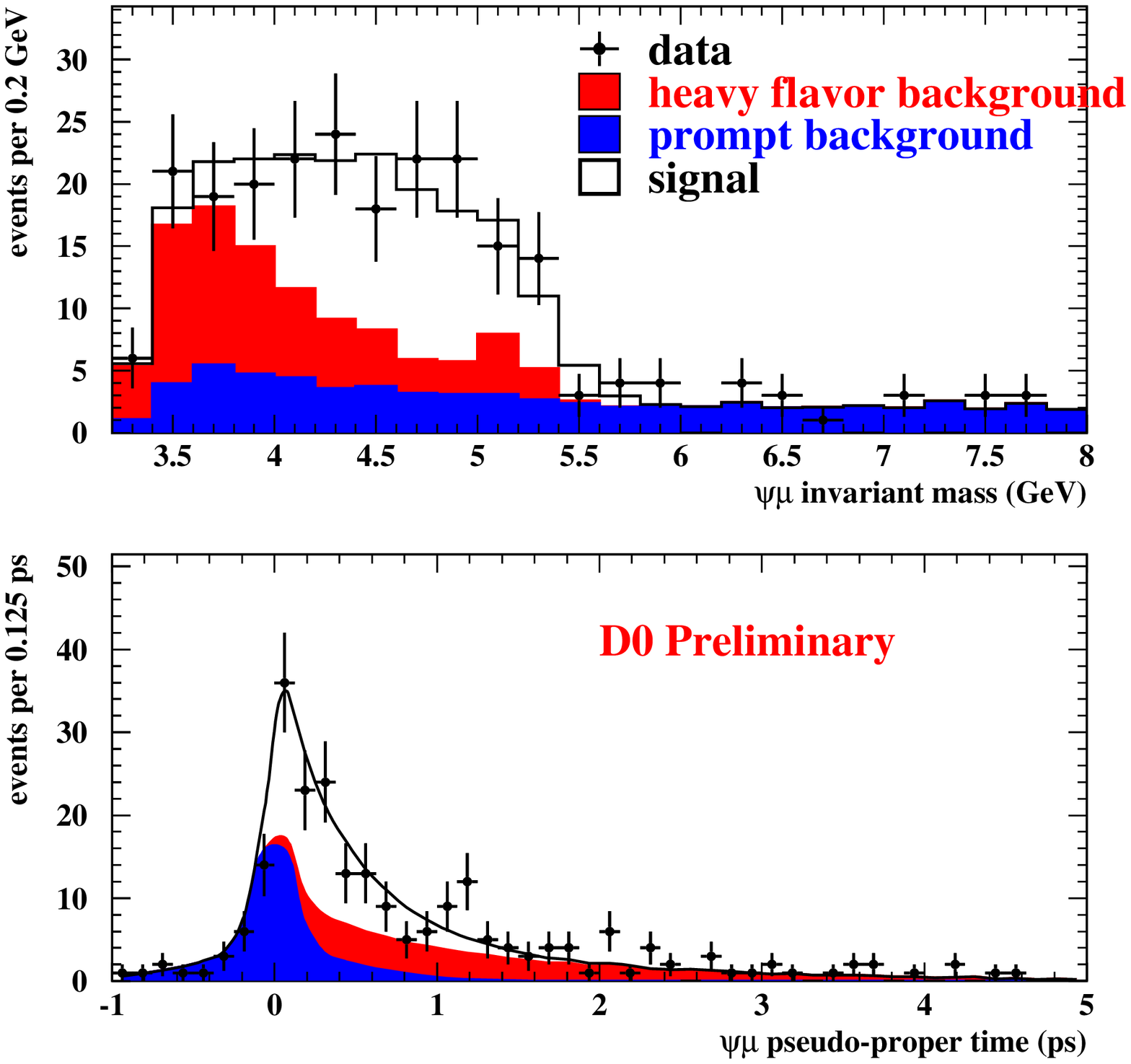,width=7cm}}
\vspace*{6pt}
\centerline{\footnotesize (a)}
\end{minipage}
\qquad\qquad\qquad
\begin{minipage}[b]{6.5cm}
\centerline{\psfig{file=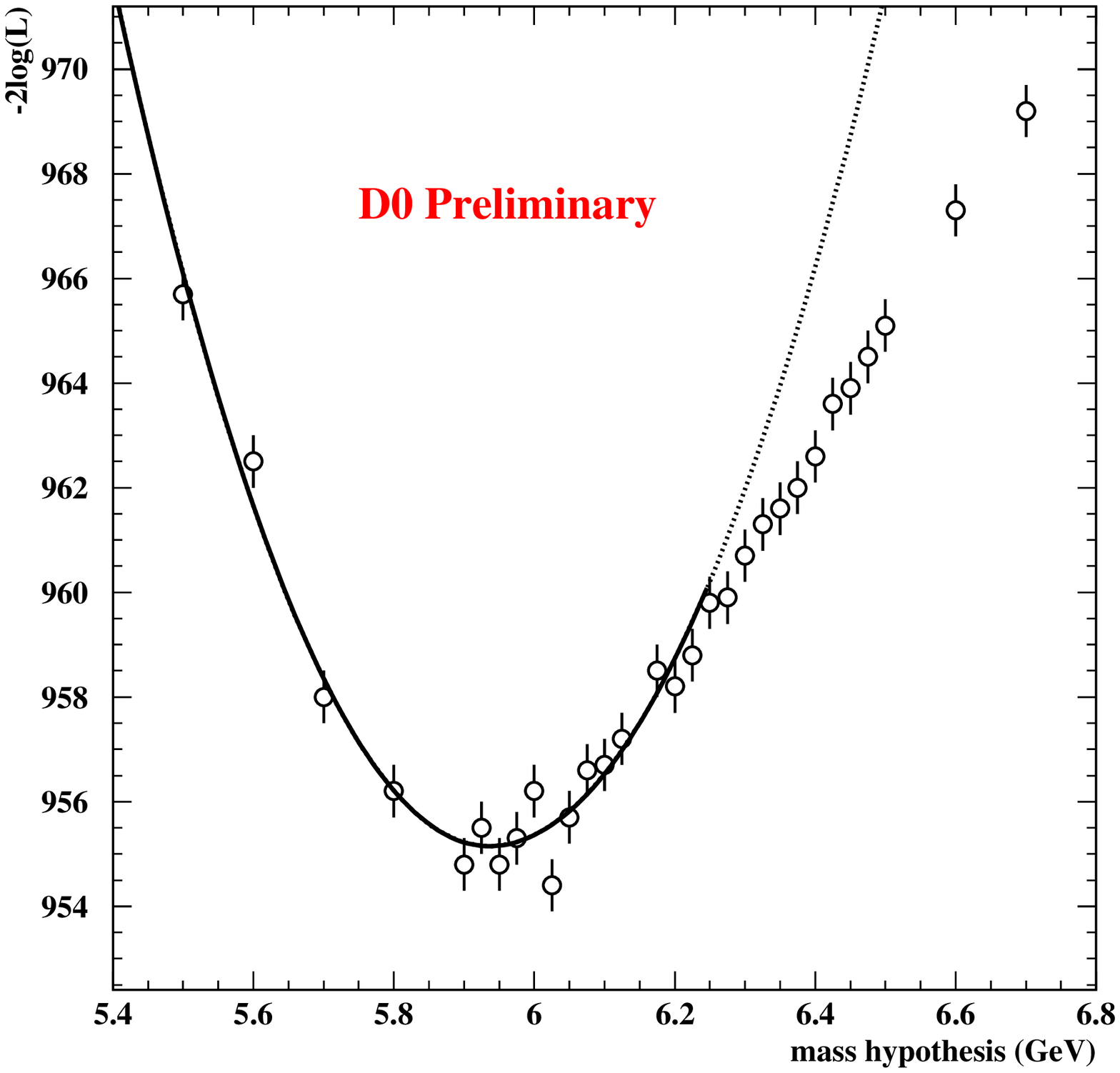,width=7cm}}
\vspace*{6pt}
\centerline{\footnotesize (b)}
\end{minipage}
}
\caption{(a) The $J/\psi\mu$ invariant mass and pseudo-proper decay
time distributions of the data candidates (points), with the results
of the best combined mass and lifetime likelihood fit overlaid, (b)
Distribution of -2log(${\cal L}$) returned by the combined fit at a
variety of mass hypotheses.\label{D0Bcplots}}
\end{figure}

We use $\sim 210$ pb$^{-1}$ of data to study this
particle\cite{D0Bc}. We combine a $J/\psi$ and a muon to form $B_c$
candidates; a background sample consisting of a $J/\psi$ and one track
(which is not a muon) is also formed. The latter sample is used to
study the heavy flavour background. Since there are missing particles
in the final state, the momentum of the $J/\psi \mu$ system must be
corrected to account for them. A simultaneous fit to the mass and
pseudo-proper decay time\footnote{This is converted to the true proper
decay time by the application of a correction factor to account for
missing particles.} of the $J/\psi$ system is performed and the
results are shown in Fig.~\ref{D0Bcplots}(a). In
Fig.~\ref{D0Bcplots}(b), we show the distribution of -2log(likelihood)
from the combined fit at a variety of mass hypotheses. A clear minimum
is observed at 5.95 GeV/c$^2$.

We observe $95\pm12\pm11$ $B_c$ candidate events, and determine the
mass and lifetime to be $5.95^{+0.14}_{-0.13}\pm0.34$ GeV/c$^2$ and
$0.448^{+0.123}_{-0.096}\pm0.121$ ps, respectively, which agrees with
the previous measurement and theoretical
predictions\cite{Bctheory,CDFBc}.

\section{Beyond Standard Model studies}

Purely leptonic decays of the B meson, {\it e.g.}, $B_s \rightarrow
\mu^+\mu^-$, are examples of Flavour Changing Neutral Currents
(FCNC). In the SM, such decays are forbidden at tree level and proceed
through the higher order box diagrams, and consequently have very low
rates.  For instance, ${\cal B}(B_s \rightarrow \mu^+\mu^-)$ is
expected\cite{Bsmumutheory} to be $(3.42\pm0.54)\times 10^{-9}$,
whereas the previous best experimental limit\cite{CDFBsmumu} is
$5.8\times 10^{-7}$ at the 90\% C.L.

Such modes are very interesting because in many BSM models, {\it e.g.}
2-Higgs Doublet, Super Symmetry, mSUGRA models, Grand Unified theories
based on SO(10) {\it etc.}, the rate can be enhanced by as much as
much as three orders of magnitude\cite{BsmumuBSM}.

\begin{figure}[hbt]
\hspace{0.5cm} 
\parbox{6.5in}{
\hfil
\begin{minipage}[b]{6.5cm}
\centerline{\psfig{file=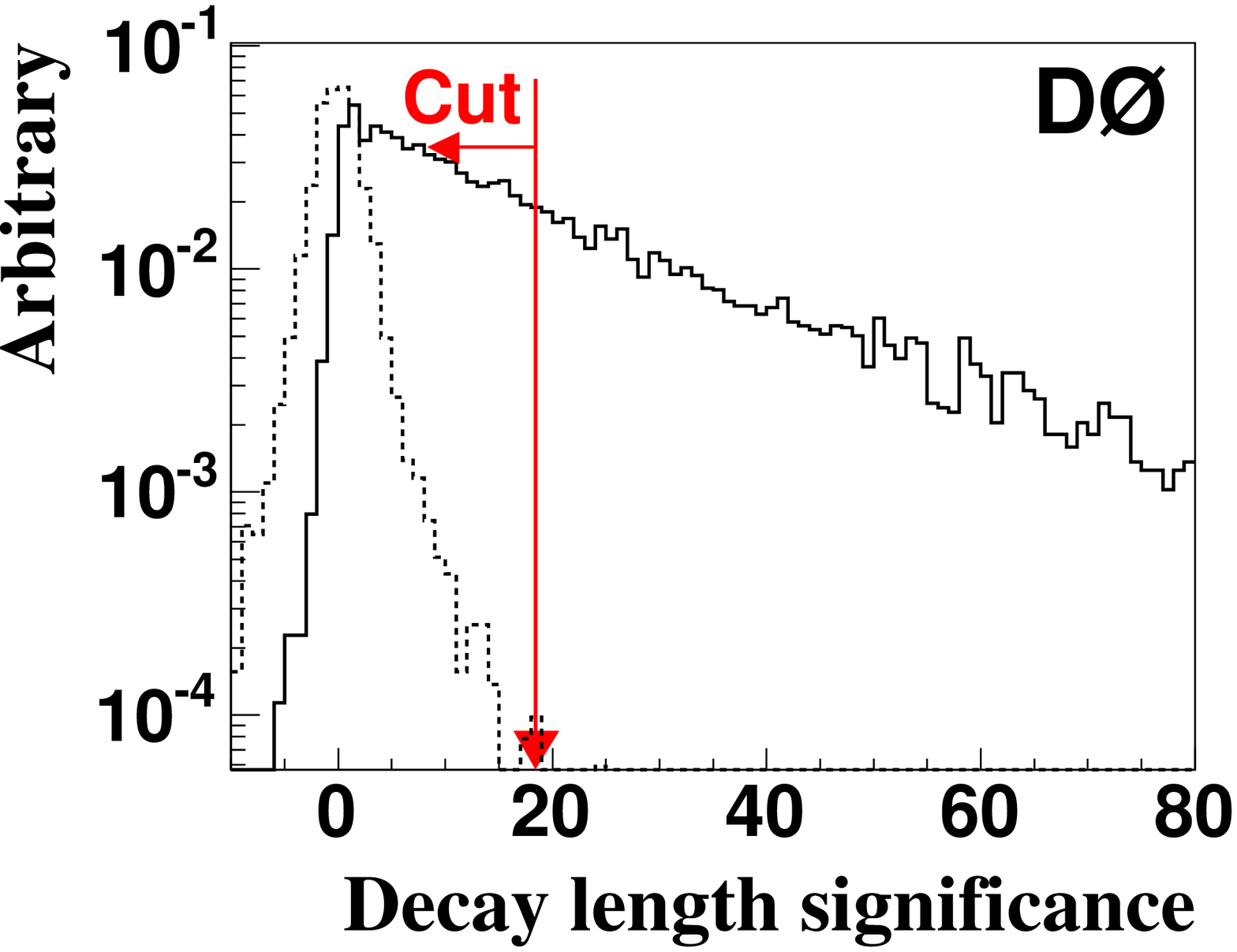,width=5.5cm}}
\vspace*{6pt}
\centerline{\footnotesize (a)}
\end{minipage}
\qquad\qquad\qquad
\begin{minipage}[b]{6.5cm}
\centerline{\psfig{file=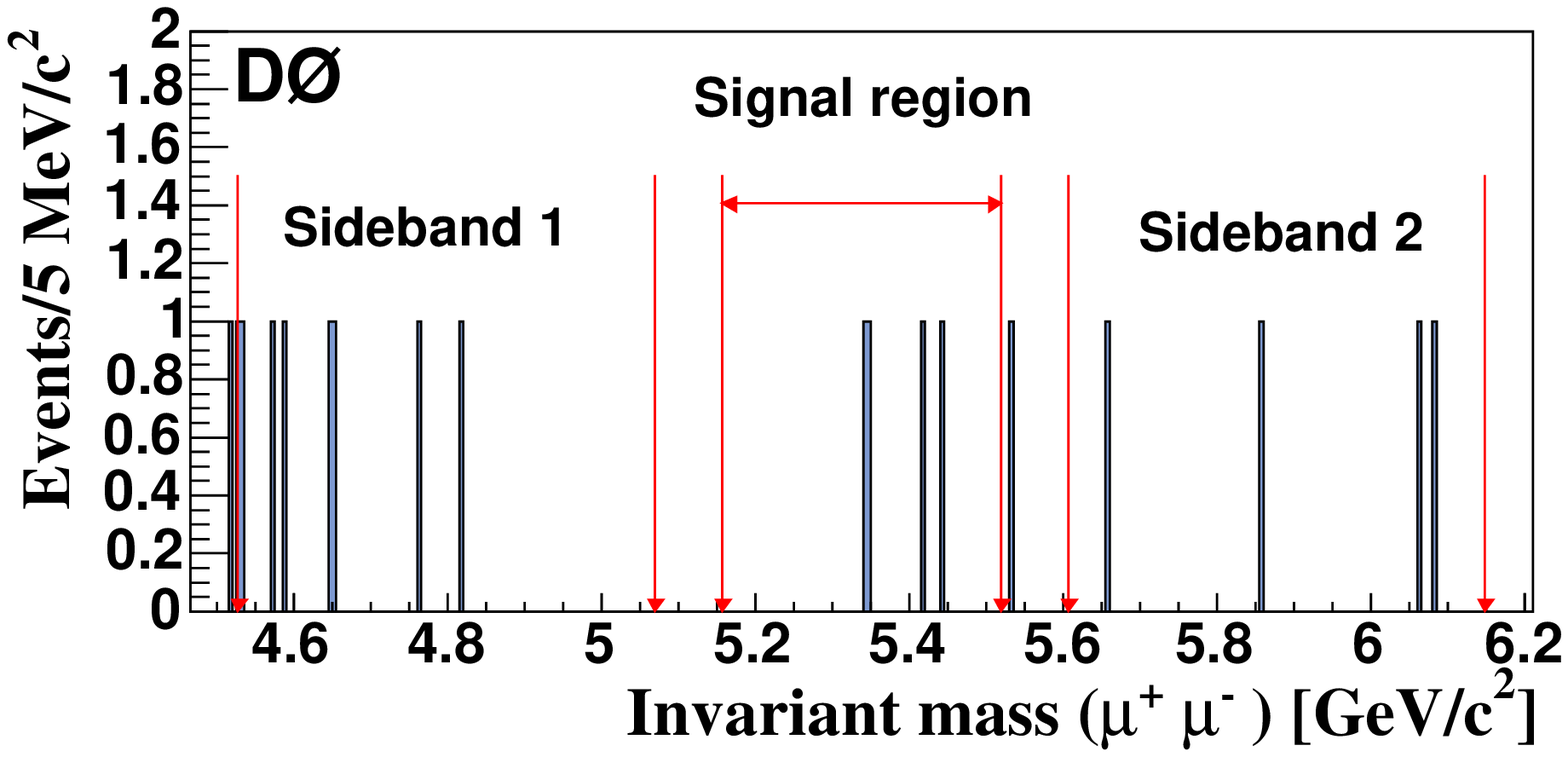,width=8.5cm}}
\vspace*{6pt}
\centerline{\footnotesize (b)}
\end{minipage}
}
  \caption{(a) ($L_{xy}/\sigma_{L_{xy}}$) after the preselection for
  signal MC (solid line) and data events (dashed line) from the
  sidebands. Arrow indicates the discriminating value that was
  obtained after optimization. Normalization is done on the number of
  signal MC and sideband data events after preselection, (b) Invariant
  mass of the remaining events of the full data sample after optimized
  requirements on the discriminating variables \label{Bsmumucuts}}
\end{figure}

We use 240 pb$^{-1}$ of data to search for this decay
mode\cite{D0Bsmumu}. After the initial (pre)selection criteria {\it
e.g.} muon quality, vertex consistency, $pT(B_s)$, {\it etc.}, we use
three additional variables to discriminate signal-like events from
background, (a) Decay Length significance ($L_{xy}/\sigma_{ L_{xy}}$)
of the $B_s$ vertex, (b) Isolation of the $\mu^+\mu^-$ pair, and (c)
Angle between the $B_s$ momentum and decay vector. These three
variables were optimized using a Monte Carlo for the signal and $B_s$
mass sidebands in data for estimating the background. In
Fig.~\ref{Bsmumucuts}(a), we show the results for one of these
variables, the optimal cut value is $L_{xy}/\sigma_{L_{xy}} > 18.5$,
as indicated by the arrow. After all (optimized) selection criteria,
the prediction is that in the ($\pm 3 \sigma$) mass region around the
$B_s$, there should be $3.7\pm1.1$ background events. With these
selection criteria, we observe 4 events, as shown in
Fig.~\ref{Bsmumucuts}(b).

To determine an upper limit, we use $B^+ \rightarrow J/\psi K^+$ as a
normalizing mode; the relative fragmentation of the b quark into $B^+$
and $B_s$ has to be taken into account. We set an upper limit,
${\cal B}(B_s \rightarrow \mu^+\mu^-) < 4.1\times 10^{-7}$ at the 90\%
C.L., which is currently the most stringent limit. 

\section{Lifetimes}

An understanding of the pattern of lifetimes of heavy hadrons provides
insight into non-perturbative QCD. In the last few years, theoretical
tools using a rigorous approach based on the heavy quark expansion (in
inverse powers of the heavy quark mass) have been
developed\cite{Bellini}. Predictions for bottom hadrons are on a much
firmer footing than for charm hadrons. Theoretical errors are further
reduced on predictions for ratios of lifetimes\cite{Tarantino}; for
instance, $\frac{\tau(B^+)}{\tau(B^0)} = 1.06\pm0.02, \;
\frac{\tau(B_s)}{\tau(B^0)}~=~1.00\pm0.01, \;
\frac{\tau(\Lambda_b)}{\tau(B^0)}~=~0.88\pm0.05$.

\subsection{Measurement of $\frac{\tau(B^+)}{\tau(B^0)}$}

Using a large sample of B semi-leptonic decays in 440 pb$^{-1}$ of
data, D\O\ has made a precision measurement of
$\frac{\tau(B^+)}{\tau(B^0)}$\cite{D0Bratioref}. B mesons were
detected via the $B \rightarrow \mu^+ \nu {\bar D^0}X$ mode (which had
$126073\pm610$ events), and categorized as the ``$D^{*-}$'' or the
``${\bar D^0}$'' sample; the former contains identified $D^{*-}
\rightarrow {\bar D^0} \pi^-$ events and is dominated by $B^0$
candidates while the latter contains the remaining events and is
dominated by $B^+$ events. The ratio of events in the two samples as a
function of proper time is primarily a function of the lifetime
difference between the $B^+$ and the $B^0$.

\vspace*{-0.2 cm}
\begin{figure}[hbt]
\hspace{0.5cm} 
\parbox{6.5in}{
\hfil
\begin{minipage}[b]{6.5cm}
\centerline{\psfig{file=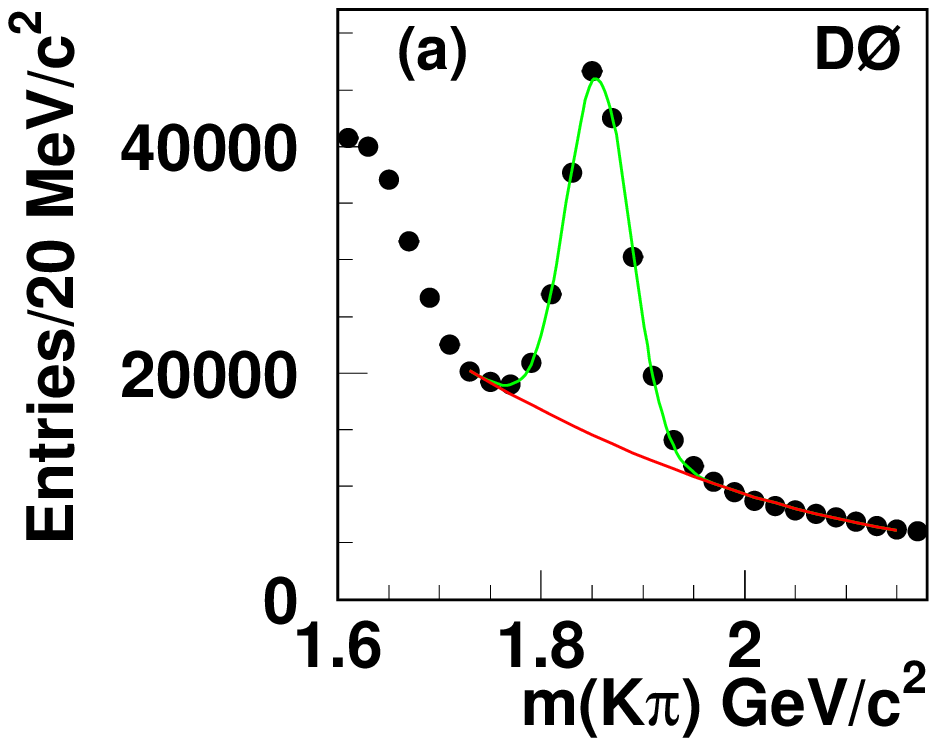,width=7cm}}
\vspace*{2pt}
\centerline{\footnotesize (a)}
\end{minipage}
\qquad\qquad\qquad
\begin{minipage}[b]{6.5cm}
\centerline{\psfig{file=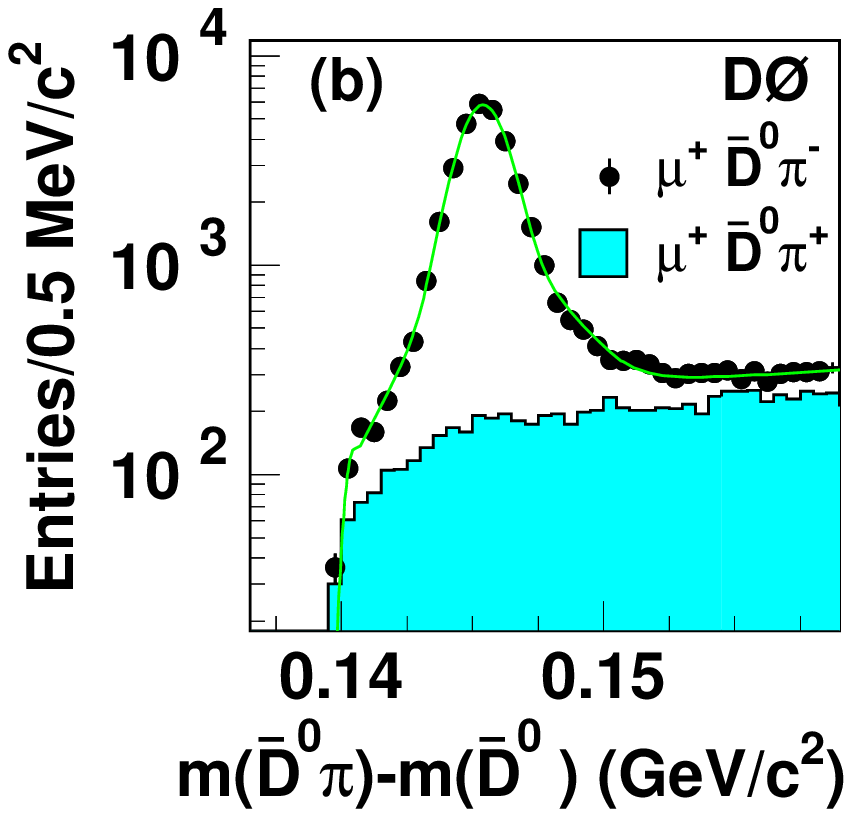,width=7cm}}
\vspace*{2pt}
\centerline{\footnotesize (b)}
\end{minipage}
}
\caption{(a)\ Invariant mass of the $K \pi$ system.  The curve shows
the result of the fit of the $K^+ \pi^-$ mass distribution. 
(b)\ Mass difference $\Delta m = m({\bar D^0} \pi) - m({\bar
D^0})$.\label{D0Bratio1}}
\end{figure}

In Fig.~\ref{D0Bratio1}, we show the invariant mass distributions
for ${\bar D^0}$ and $D^{*-}$ candidates, where they have been
combined with a muon of the correct charge, {\it e.g.,} $B^+
\rightarrow {\bar D^{(*)} } \mu^+ X$. Fig.~\ref{D0Bratio1}(b) also
shows the wrong-sign spectrum; as expected there is no peak.

Candidate events are classified as $D^0$ and right (and wrong) sign
$D^*$. From these categories, we determine the number of true ${\bar
D^0} \mu^+$ and ${D^{*-} } \mu^+$ events in eight bins of visible
proper decay length (VPDL), where $VPDL = m_B c \left( \mbox{
$L_T$} \cdot \mbox{$p$}_T(\mu^+ {\bar D^0}) \right) /
|\mbox{$p$}_T(\mu^+ {\bar D^0})|^2$, and calculate the
ratio, $r_i = \frac{ N_i({D^{*-} } \mu^+)}{N_i( {\bar D^0}\mu)}$. {\bf
$L_T$} is the transverse decay length of the $(\mu^+ {\bar D^0})$
vertex relative to the primary vertex, and {\bf $p_T$} is the
transverse momentum. To avoid any biases, for both ${\bar D^0}$ and
$D^{*-}$ samples, the soft pion is not used for determining VPDL or
$p_T$.

Next we determine the expected ratio of events in each VPDL bin and
minimize, $\chi^2 = \sum_i \frac{(r_i -
r_i^e(\epsilon_\pi,k))^2}{\sigma^2(r_i)}$, where the lifetime ratio,
$k = \tau^+/\tau^0 - 1$, and the efficiency of the slow pion
$\epsilon_\pi$ are free parameters. The $r_i^e$'s are determined {\it
ab initio} using $B$ semi-leptonic branching fractions, detector
resolution, a Monte Carlo simulation to account for missing particles,
reconstruction efficiencies, and the world average for the $B^+$
lifetime; $B^0$ lifetime, $\tau^0$, is expressed as $\tau^0 =
\tau^+/(1+k)$.

\begin{figure}[th]
\centerline{\psfig{file=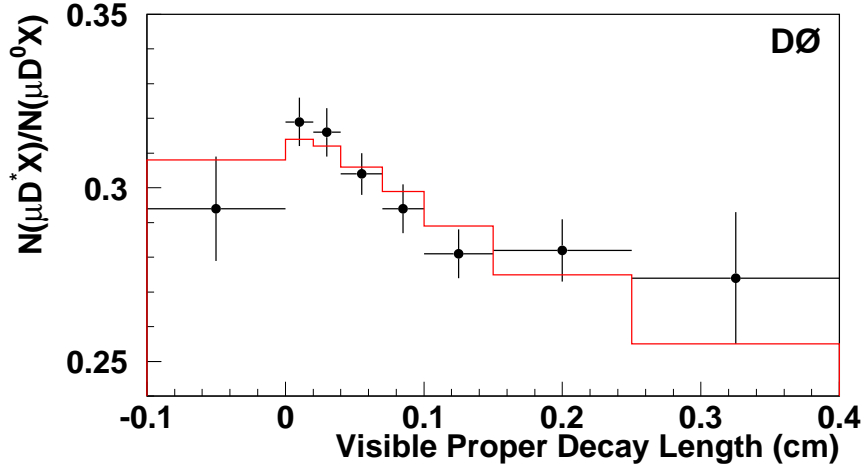,width=5in}}
\vspace*{4pt}
\caption{Points with the error bars show the ratio of the number 
of events in the $\mu^+ D^{*-}$ and $\mu^+ {\bar D^0}$ samples
as a function of the visible proper decay length. The result
of the minimization with $k=0.080$ is shown as a histogram.\label{D0Bratio2}}
\end{figure}

Minimization of the $\chi^2$ distribution gives, $ k =
\tau^+/\tau^0 - 1 = 0.080 \pm 0.016 (stat) \pm 0.014 (syst)$. In
Fig.~\ref{D0Bratio2}, we show $r_i$ as a function of VPDL; the fit
results are overlaid. The systematic uncertainty includes
uncertainties on $B$ branching fractions, reconstruction efficiencies,
detector resolutions, {\it etc.}

\subsection{Measurement of $\tau(\Lambda_b)$}

Theoretical predictions for the ratio,
$\frac{\tau(\Lambda_b)}{\tau(B^0)}$ ($\sim (0.88-0.98)$ with errors
$\sim \pm 0.02-0.05$)\cite{LambdaBtheory}, have always been somewhat
higher than experimental results\cite{PDG}, ($0.80 \pm 0.05$). One
possible source for the discrepancy could be that all previous
measurements used partially reconstructed $\Lambda_b$, which required
one to use Monte Carlo simulations to correct for missing particles,
{\it e.g.,} $\nu, \pi$ {\it etc.}. Since $\Lambda_b$ decays have not
been studied in detail, this correction process makes the result model
dependent.

We measure $\tau(\Lambda_b)$, via the fully reconstructed decay,
$J/\psi \Lambda$, with $J/\psi \rightarrow \mu^+\mu^-, \; \Lambda
\rightarrow p\pi^-$, using $\sim 250$ pb$^{-1}$ of integrated
luminosity. For comparison, $\tau(B^0)$ is measured using the $J/\psi
K_s^0$ mode ($K_s^0 \rightarrow \pi^+\pi^-$); the two modes have
similar topology.  We reconstruct $61\pm 12$ and $291\pm 23$ events,
respectively\cite{D0LambdaB}.

\begin{figure}[hbt]
\hspace{0.5cm}
\parbox{6.5in}{
\hfil
\begin{minipage}[b]{6.5cm}
\centerline{\psfig{file=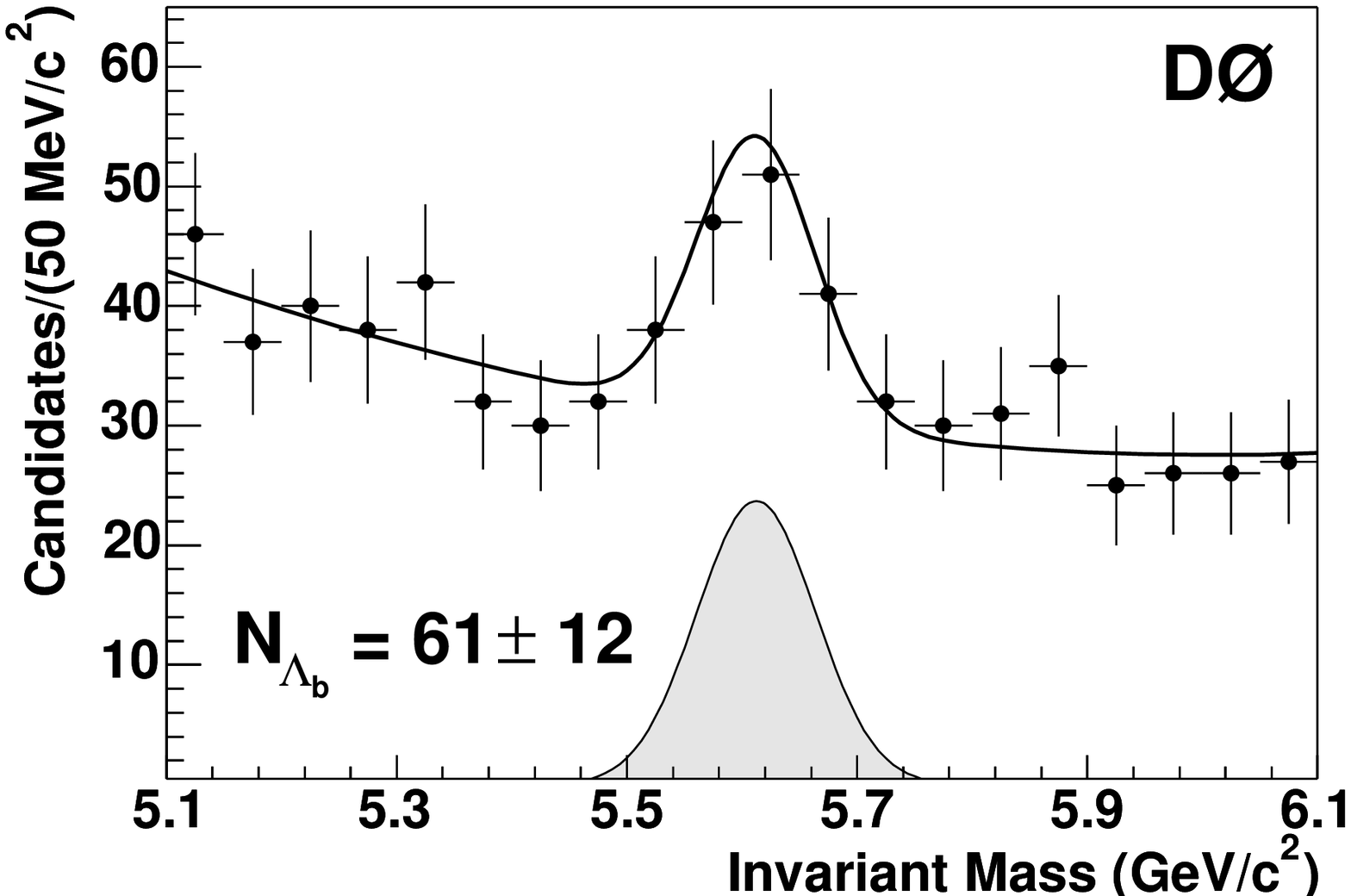,width=7cm}}
\vspace*{2pt}
\centerline{\footnotesize (a)}
\end{minipage}
\qquad\qquad\qquad
\begin{minipage}[b]{6.5cm}
\centerline{\psfig{file=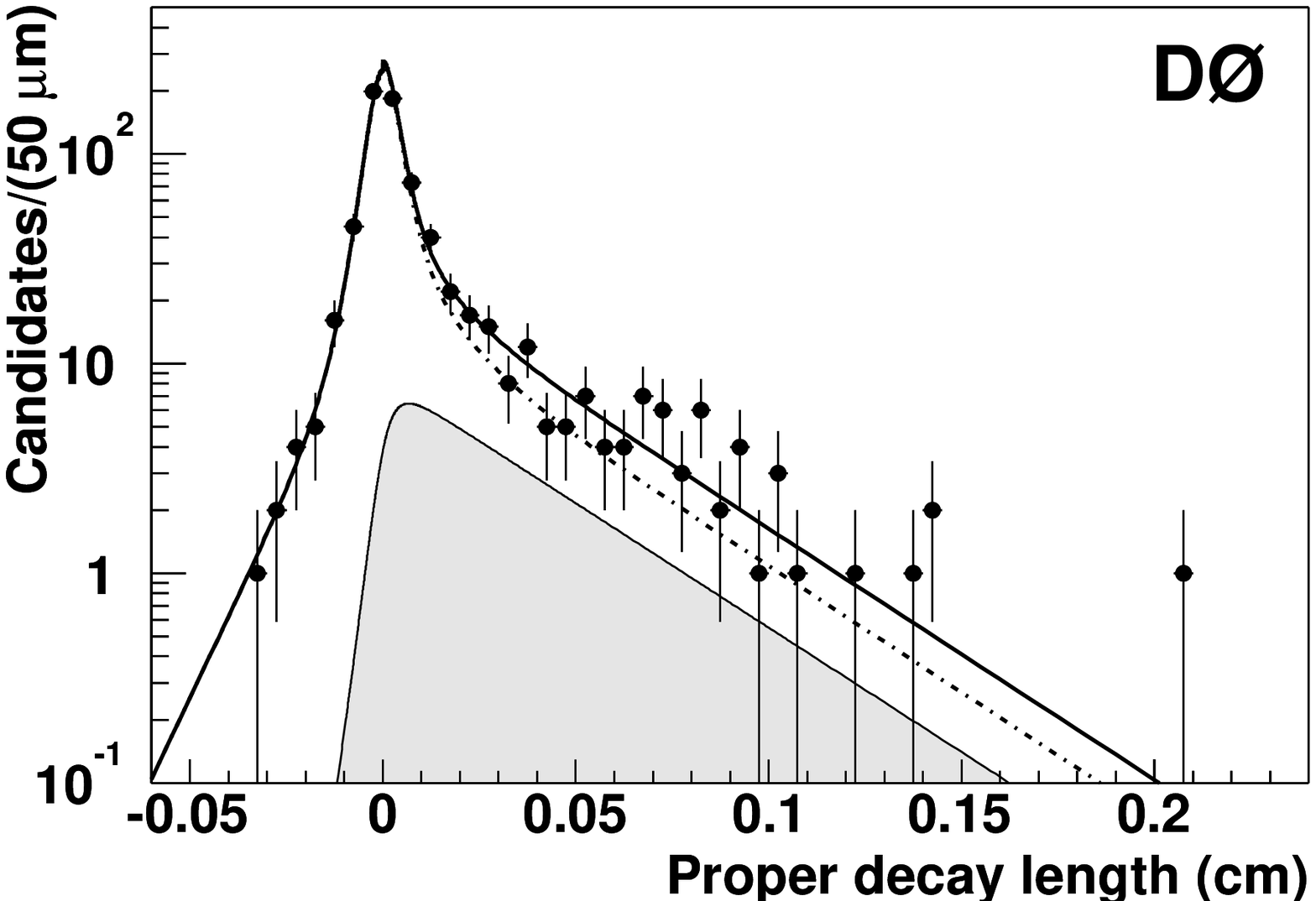,width=7cm}}
\vspace*{2pt}
\centerline{\footnotesize (b)}
\end{minipage}
}
\caption{(a)Invariant mass distribution for $\Lambda^0_b$ candidate
events.  The points represent the data, and the curve represents the
result of the fit.  The mass distribution for the signal is shown in
gray, (b) Distribution of proper decay length for $\Lambda^0_b$
candidates. The points are the data, and the solid curve is the sum of
the contributions from signal (gray) and the background (dashed-dotted
line) \label{Lambdaplots}}
\end{figure}

In Fig.~\ref{Lambdaplots}(a), we show the mass distribution of the
$\Lambda^0_b$ candidate events, and in Fig.~\ref{Lambdaplots}(b) their
proper decay lengths are presented. 

The lifetime was obtained by performing a unbinned likelihood fit
using both the mass and the lifetime of candidate events. We used all
events in the mass range 5.1-6.1 GeV/c$^2$ for $\Lambda_b$ (4.9-5.7
GeV/c$^2$ for $B^0$). Different functions were used to model the
lifetime (and mass) of signal and background events.

We obtain, $\tau(\Lambda^0_b) = 1.22^{+0.22}_{-0.18} \mbox{ (stat)}
\pm 0.04 \mbox{ (syst) ps,}$ and $ \tau(B^0) = 1.40^{+0.11}_{-0.10}
\mbox{ (stat)} \pm 0.03 \mbox{ (syst) ps}$.  This is the first time
that the $\Lambda_b$ lifetime has been measured in an exclusive
channel; our result agrees with previous measurements\cite{PDG}. We
also obtain, $\frac{\tau(\Lambda^0_{b})}{\tau(B^0)} =
0.87^{+0.17}_{-0.14} \mbox{ (stat)}\pm 0.03 \mbox{ (syst)}$.  The
current statistical error on the ratio is too large to draw any
conclusions, however this error will decrease as more data is
processed.  The systematic uncertainty includes contributions from
Silicon alignment uncertainties, models used for signal and background
and cross-feed between the $J/\psi \Lambda$ and $J/\psi K^0_s$ modes.

\subsection{Measurement of $\tau(B_s^0)$}

Theoretical calculations predict $\frac{\tau(B_s^0)}{\tau(B^0)} \sim
1.0 \pm {\cal O}(1\%)$\cite{Bellini,Tarantino}. The CP
eigenstates of the $B_s^0-{\bar B_s^0}$ system are expected to have
different lifetimes\cite{BsCPtheory}; predictions for $\frac{\Delta
  {\Gamma_s}}{\Gamma_s}$ are $\sim 10-20\%$. This difference can be
probed by comparing the $B_s$ lifetime measured using the
semi-leptonic final state (which has equal mixtures of the two CP
states) and using a decay like $J/\psi\phi$ (which is expected to be
dominantly CP even). Alternatively, one could directly measure the
lifetime difference using the $J/\psi\phi$ mode; this analysis has
recently been performed by the CDF collaboration and they find
$\frac{\Delta{\Gamma_s}}{\Gamma_s} = 65^{+25}_{-33}\pm
1\%$\cite{CDFBsCP}.

Using $\sim 220$ pb$^{-1}$ of integrated luminosity, D\O\ has measured
the $B_s^0$ lifetime reconstructed in the $J/\psi\phi \; (\phi
\rightarrow K^+K^-)$ final state\cite{D0Bslife}. In this analysis, we
fit the $B_s^0$ lifetime with a single exponential, {\it i.e.,} under
the assumption $\Delta \Gamma_s \sim 0$.  For comparison, we also
measure $B^0$ lifetime using the final state $J/\psi K^{*0} \; (K^{*0}
\rightarrow K^+\pi^-)$, which has similar topology; the two modes have
337 and 1370 signal events, respectively.

\begin{figure}[hbt]
\hspace{0.5cm}
\parbox{6.5in}{
\hfil
\begin{minipage}[b]{6.5cm}
\centerline{\psfig{file=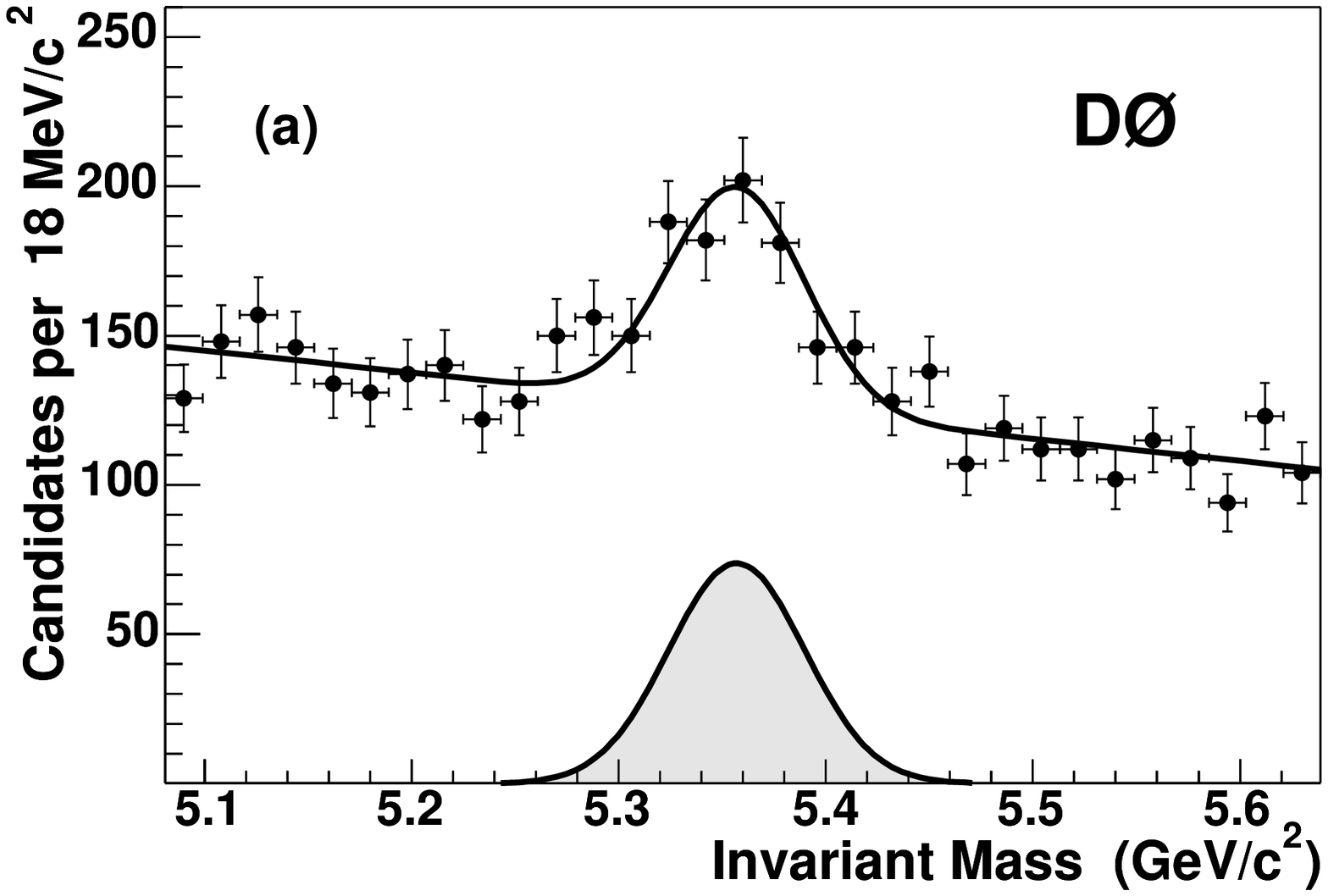,width=7cm}}
\vspace*{2pt}
\centerline{\footnotesize (a)}
\end{minipage}
\qquad\qquad\qquad
\begin{minipage}[b]{6.5cm}
\centerline{\psfig{file=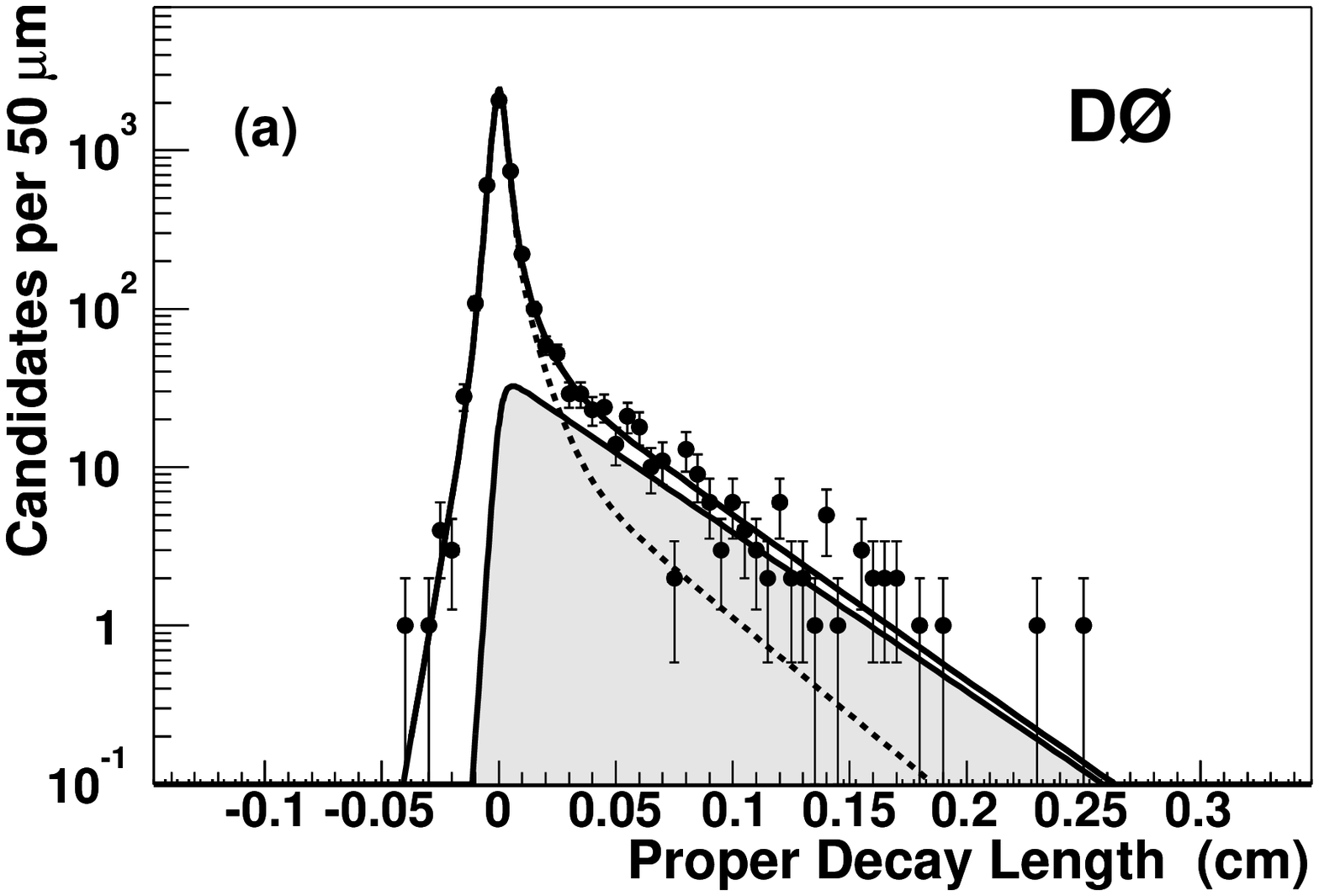,width=7cm}}
\vspace*{2pt}
\centerline{\footnotesize (b)}
\end{minipage}
}
\caption{(a)Invariant mass distribution for $B_s^0$ candidate
events.  The points represent the data, and the curve represents the
result of the fit.  The mass distribution for the signal is shown in
gray, (b) Distribution of proper decay length for $B_s^0$
candidates. The points are the data, and the solid curve is the sum of
the contributions from signal (gray) and the background (dashed line) 
\label{Bsplots}}
\end{figure}

In Fig.~\ref{Bsplots}(a), we show the mass distribution of the $B_s^0$
candidate events, and in Fig.~\ref{Bsplots}(b) their proper decay
lengths are presented.  The measurement technique is the same as the
one used in the $\Lambda_b$ lifetime analysis.

We find $\tau(B_{s}^0) = 1.444 ^{+0.098}_{-0.090}\ (\mbox{stat}) ~ \pm
0.020\ (\mbox{sys})\ \mbox{ps,}$ and $\tau(B^0) = 1.473
^{+0.052}_{-0.050}\ (\mbox{stat}) \pm 0.023\ (\mbox{sys})\ \mbox{ps.}$
Both results are consistent with world averages.\footnote{Since the
  $B_s$ average ($1.461\pm0.057\;\mbox{ps}$)\cite{PDG} is dominated by
  measurements made using semi-leptonic final states, our result 
  appears to prefer a low value for $\frac{\Delta
    \Gamma_s}{\Gamma_s}$.} In addition, we also measure,
$\frac{\tau(B_{s}^0) }{\tau(B^0) } = 0.980\ ^{+0.075}_{-0.070}\
(\mbox{stat}) \ \pm 0.003\ (\mbox{syst})$, which is in agreement with
theoretical predictions.

\section{$B_s{\bar B_s}$ Mixing}

The phenomenon of particle-antiparticle mixing has yielded many
unexpected results and has provided the impetus for significant
progress in the field. For instance, the large rate of mixing in
$B^0{\bar B^0}$ implied that the top quark was much heavier than
previously expected and $K^0-{\bar K^0}$ mixing taught us about CP
violation. Quark mixing occurs at the one-loop level via ``box''
diagrams and heavy particles (in the loop) tend to have enormous
influence\cite{Buras}.

The study of $B_s{\bar B_s}$ mixing has a twofold purpose. Given the
previous successes of mixing induced processes, one could hope for a
surprise. Failing an unexpected result, the measurement of the rate of
$B_s{\bar B_s}$ mixing will aid in reducing the error on the
measurement of the CKM element $V_{td}$\cite{PDG}. If our current
understanding of the Standard Model and the CKM matrix are
correct\cite{CERN:yellow}, then \Bs oscillations should occur with a
frequency, $\Delta M_s$, in the (95\% CL) interval (14.2-28.1)
ps$^{-1}$. A deviation could be a sign of new physics.  Additionally,
if the mixing parameter, $\Delta M_s$ is very large then the
difference in the widths of the CP eigenstates of the \Bs may be
detectable.\footnote{This analysis is in progress.}

B oscillations are observed by comparing the proper time evolution of
events where a neutral B-hadron decays as a particle of the opposite
flavor from that with which it was produced (mixed B) to those where
the B-hadron's production and decay flavors are the same (unmixed B).
To study \Bs oscillations we therefore need three ingredients, (a)
final state reconstruction, (b) ability to measure $B$ decay lengths,
and (c) flavour tagging of the $B$ both at production and decay.

The significance of a $B$ mixing measurement can be expressed as,
\begin{equation}
{\rm Sig} = \sqrt{\frac{N\epsilon D^2}{2}}\exp^{-(\Delta M\times \sigma_t)^2/2}\sqrt{\frac{S}{S+B}}
\label{signif}
\end{equation}

where $N$ is the number of reconstructed \Bs events, $\epsilon D^2$ is
a measure of how well we know the flavour of the \Bs at production;
$\epsilon$ is the efficiency of the tag, and $D$, the dilution is
defined as $D=1 -2w$, where $w$ is the probability of mis-identifying
the flavour of the B-hadron at production.  $\sigma_t$ is the proper
time resolution, and $S/(S+B)$ expresses signal purity. It is clear
that as $\Delta M_s$ gets larger, the importance of good proper time
resolution increases.

At D0, B-mixing is studied mainly using semileptonic $B^0$ and $B^0_s$ decays,
although a fully hadronic Bs decay mode analysis is in progress. The
advantage of using semi-leptonic events vis-a-vis (fully
reconstructed) hadronic events is that the branching fraction for the
former is much larger than for the latter; the total rate for $B_s
\rightarrow D_s\mu\nu + X$ is $\sim 10\%$, whereas the rate for $B_s
\rightarrow D_s^{(*)-}\pi^+$ is $\sim 0.3\%$. The disadvantage of
using semi-leptonic is that due to missing particles, the proper time
resolution is worse.

\begin{figure}[hbt]
\hspace{0.5cm}
\parbox{6.5in}{
\hfil
\begin{minipage}[b]{6.5cm}
\centerline{\psfig{file=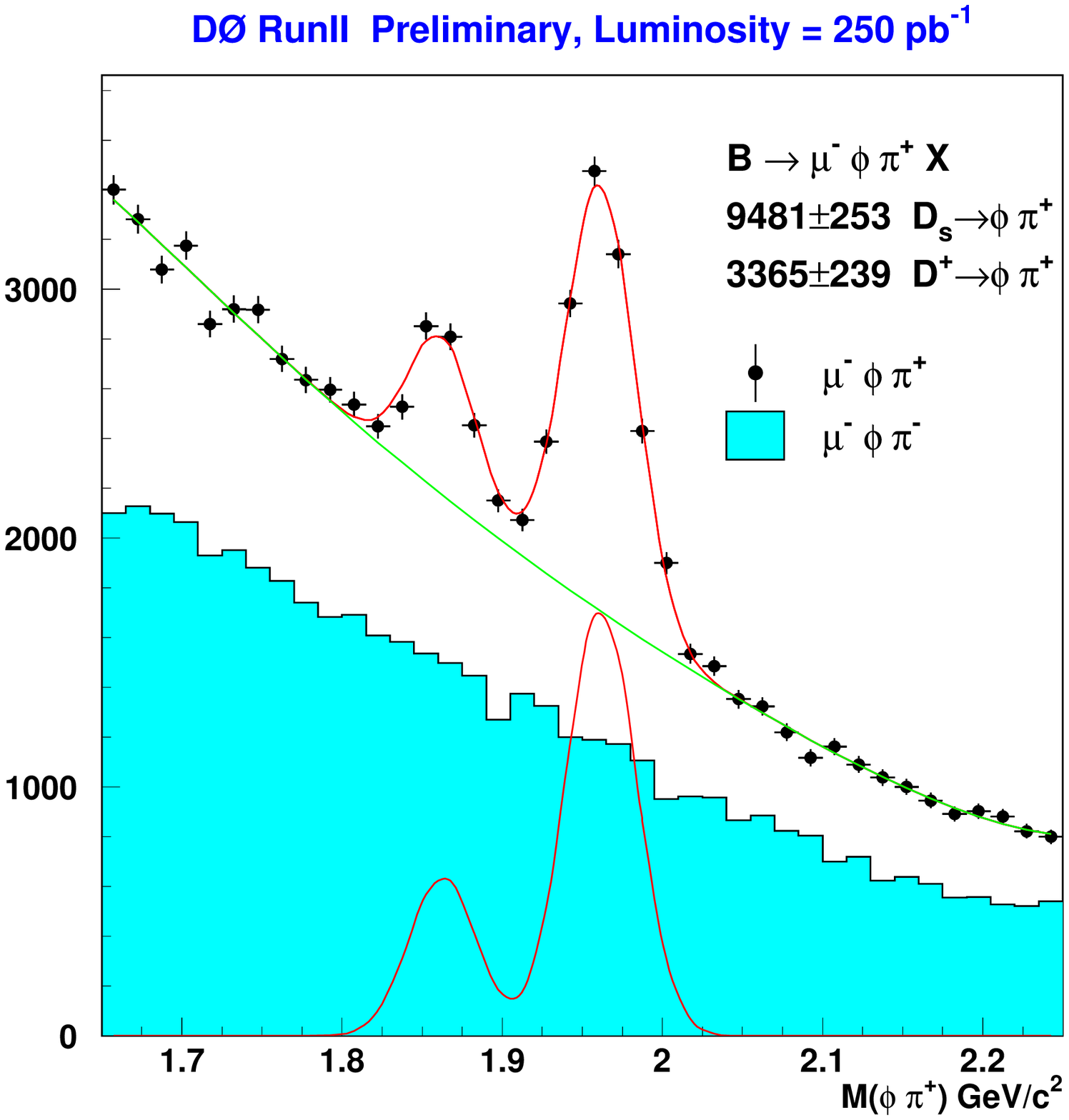,width=7cm}}
\vspace*{2pt}
\centerline{\footnotesize (a)}
\end{minipage}
\qquad\qquad\qquad
\begin{minipage}[b]{6.5cm}
\centerline{\psfig{file=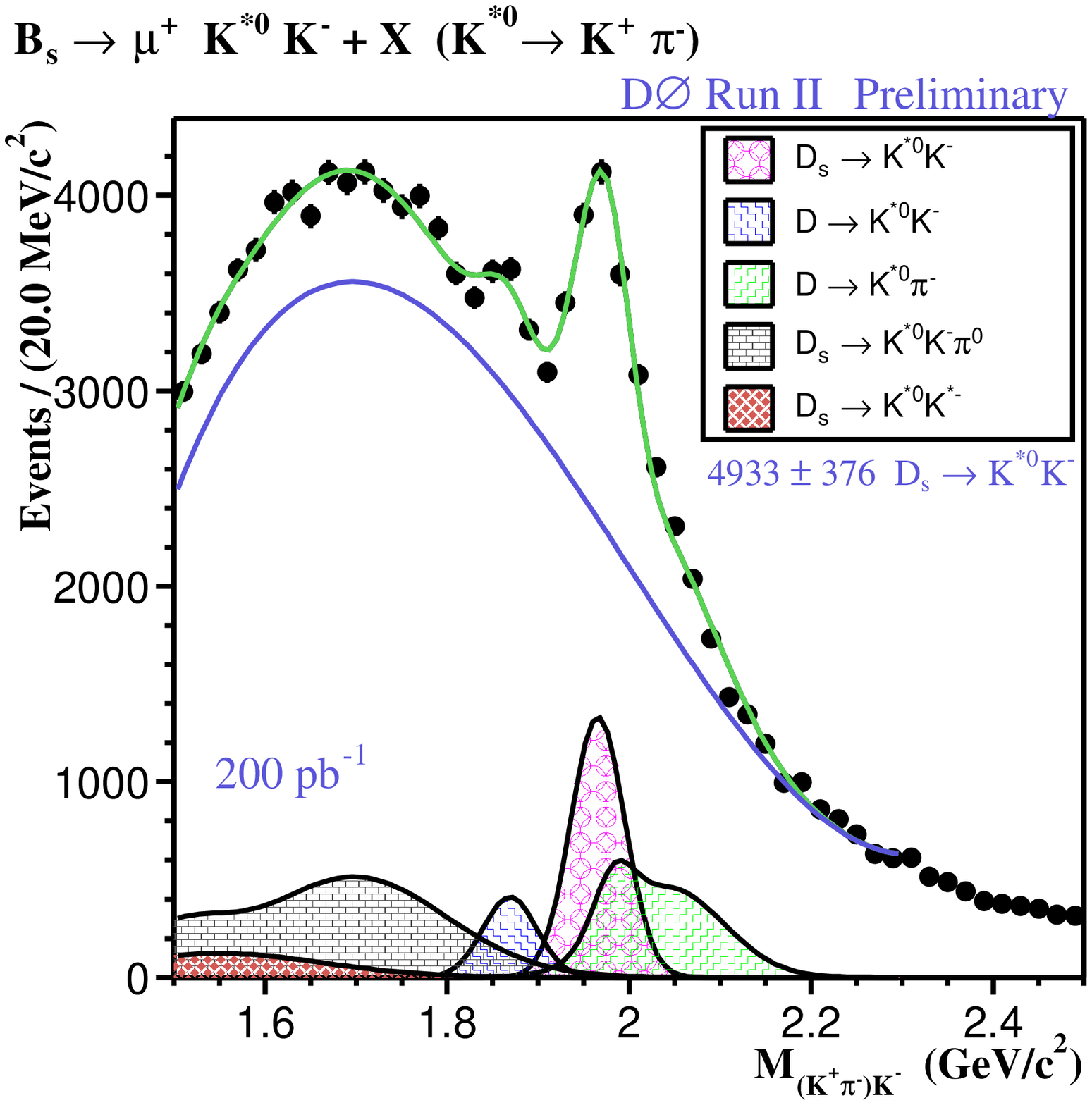,,width=7cm}}
\vspace*{2pt}
\centerline{\footnotesize (b)}
\end{minipage}
}
\caption{Invariant mass distribution for $D_s^{\pm}$ candidate events
where the $D_s$ is charge correlated to the muon, such that the peaks
correspond to right sign $B_s \rightarrow D_s\mu\nu + X$ events for
(a) $D_s^+ \rightarrow \phi\pi^+$, (b) $D_s^- \rightarrow K^{*0}K^-$.  
The blue shaded histogram in (a) corresponds to wrong
sign $D_s \mu$ combinations.  The peaks in (b) correspond to $D_s$
($D^+$) signal and background. \label{BsSL}}
\end{figure}

In Fig.~\ref{BsSL}, we present the inclusive $B \rightarrow D_s \mu X$
signal for (a) $D_s^+ \rightarrow \phi\pi^+$ using 250 $pb^{-1}$ and
(b) $D_s^- \rightarrow K^{*0}K^-$ using 200 $pb^{-1}$. The peaks at
the $D_s$ mass are dominated by $B_s$ decays.

The next component needed for a mixing measurement is knowledge of the
flavour of the $B$ hadron at the time of production and decay. By
using flavour-specific decays, one can easily tag the flavour at the
time of decay. To tag the $B$ flavour at production we use the
following techniques,

\begin{itemize}
\item
   Soft Lepton Tagging: The sign of the lepton produced in the
   semi-leptonic decay of the other $B$ in the event is used to tag
   the flavour of the signal $B$. We then make the assumption that (at
   production) the flavour of the signal $B$ is opposite to that of
   the tag $B$.  This method has low efficiency, but very high tagging
   power. We are also using electrons, although those results are not
   final as yet.

\item
   Jet Charge Tagging: We take all tracks opposite to the signal $B$
   and form a track jet, and measure its charge.  The assumption is
   that these tracks are produced in the fragmentation of the other
   b-quark, as well as in the decay of the tag $B$ hadron.  This
   method has high efficiency, but has poorer tagging power.

\item 
   Same Side Tagging: In this technique, we identify particles
   produced in the fragmentation of the b-quark which gives rise to
   the signal $B$. In addition, the signal B can come from a resonance,
   {\it e.g.,} B$^{**+} \rightarrow B^0\pi^+$, and the charge of such
   pions is correlated with the flavour of the signal B at the time
   of production. This method has high efficiency, but has poorer
   tagging power.

\end{itemize} 

\begin{figure}[hbt]
\hspace{0.5cm}
\parbox{6.5in}{
\hfil
\begin{minipage}[b]{6.5cm}
\centerline{\psfig{file=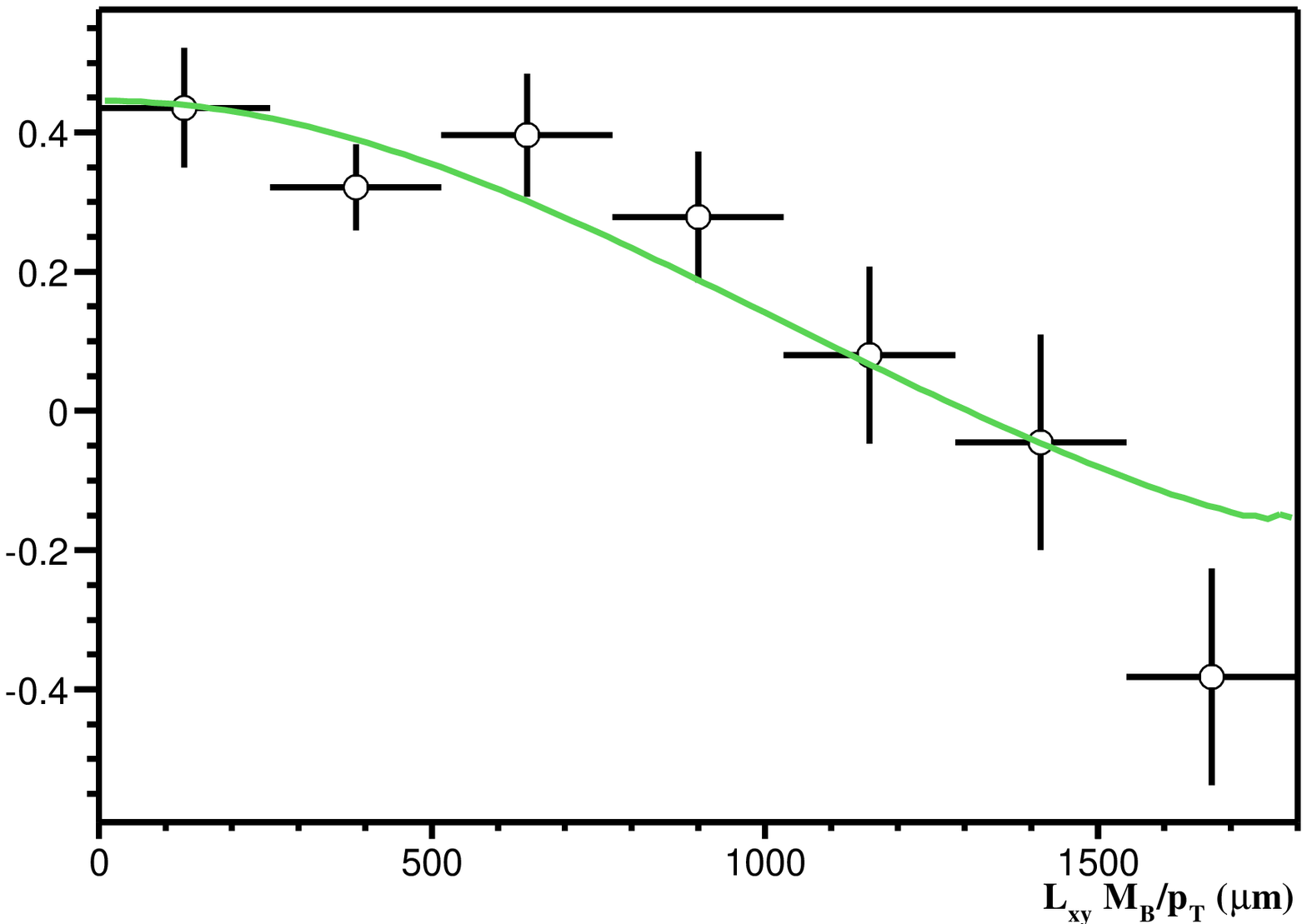,width=7cm}}
\vspace*{2pt}
\centerline{\footnotesize (a)}
\end{minipage}
\qquad\qquad\qquad
\begin{minipage}[b]{6.5cm}
\centerline{\psfig{file=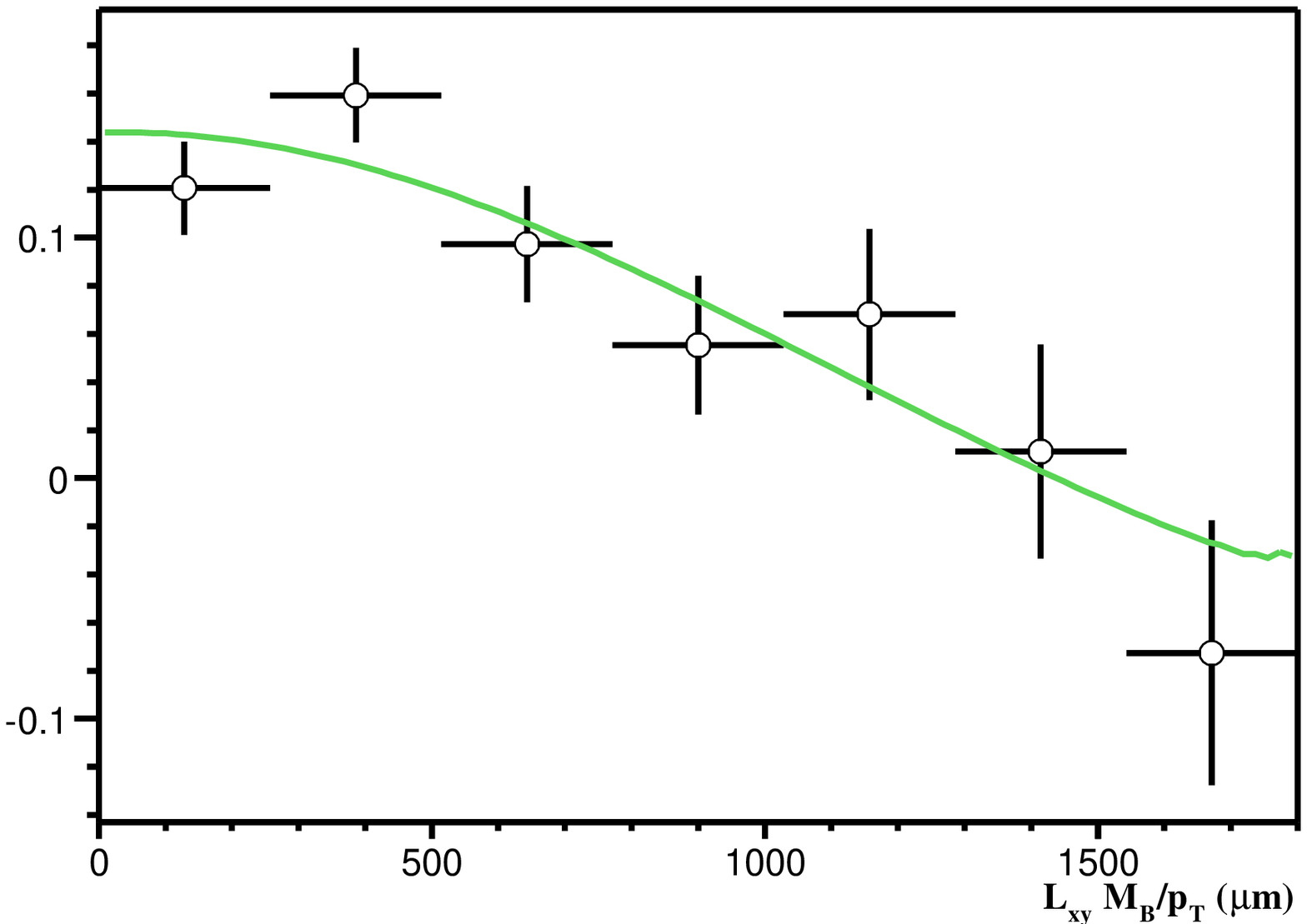,width=7cm}}
\vspace*{2pt}
\centerline{\footnotesize (b)}
\end{minipage}
}
\caption{Asymmetry distributions for $B^0 \rightarrow D^{*-}\mu^+\nu
+ X$ events, using (a) the soft muon tag, (b) a combination of jet
charge and same side tags. \label{Bdmixing}}
\end{figure}

We have tested our analysis by measuring the mixing parameter for
$B^0{\bar B^0}$, {\it i.e.,} $\Delta M_d$; we use $\sim 200$ pb$^{-1}$
for this study.  In Fig.~\ref{Bdmixing}, we show the asymmetry as a
function of visible proper decay length, where the asymmetry, ${\cal
A}(t)$, is defined as $\frac{N_U - N_M}{N_U + N_M}$; $N_U (N_M)$ are
the numbers of unmixed (mixed) events in the various time bins. In
this analysis, we use the decay mode $B^0 \rightarrow
D^{*-}\mu^+\nu+X$ events.  In (a), we show the asymmetry for events
tagged with a soft muon, whereas in (b) we show the asymmetry for
events tagged with a combination of Jet Charge and Same Side taggers.
A simultaneous fit to the two distributions in Fig.~\ref{Bdmixing}
yields $\Delta M_d = 0.456 \pm 0.034 ({\rm stat}) \pm 0.025 ({\rm
syst})$, which is in agreement with the world average\cite{PDG}. We
also get the efficiency and dilution for the two tags to be, (a)
$\epsilon = 5.0\pm0.2\%$ and $D = 44.8\pm5.1 \%$, (b) $\epsilon =
68.3\pm0.9\%$ and $D = 14.9\pm1.5 \%$, respectively. 

Since $\Delta M_s$ is much larger than $\Delta M_d$, it is clear from
Eq.~(\ref{signif}) that the critical element is the proper time
resolution. Work is underway to improve the resolution for
semi-leptonic final states.

We also have other improvements in the pipeline that will have
significant impact on the $\Delta M_s$ analysis: (a) A new layer of
silicon sensors will be installed at a radius of 2.5 cm. This is
expected to improve the proper time resolution by about 25\%, (b) An
increase in the trigger bandwidth will enable us to increase the yield
of ``B-rich'' events many-fold, and (c) Improved triggers will enrich
the data with hadronic $B_s$ decays.

\section{Conclusions}

The D0 detector has started to produce very competitive results in the
field of $B$ physics. We have already recorded $\sim 520$ pb$^{-1}$ of
data and hope to collect $\sim 1 fb^{-1}$ by the end of 2005 and $\sim
4 (8) fb^{-1}$ by the end of 2007 (2009).

As a stepping stone to \Bs mixing, we have measured $\Delta M_d$ to
benchmark our analyses techniques. In addition, we are pursuing a
vigorous program which includes measurement of $B$ lifetimes, rare
decays, studies of quarkonia\cite{Upsilon}, beauty baryons and $B_c$.

\section*{Acknowledgments}

I would like to thank Hal Evans, Brad Abbott and Jesse Ernst for stimulating discussions.

\end{document}